\documentclass[useAMS, twocolumn, usenatbib]{mn2e}
\usepackage{graphicx}
\usepackage{amsmath, amssymb}
\usepackage{subfigure}
\usepackage{natbib}

\newcommand       \apj          {ApJ}
\newcommand       \apjl         {ApJL}
\newcommand       \aap          {A\&A}
\newcommand       \nat          {Nature}
\newcommand       \mnras        {MNRAS}

\newcommand       \aj      {AJ}
\newcommand       \prd      {Phys.~Rev.~D.~}

\newcommand       \araa      {ARA\&A}

\newcommand      \apjs {ApJ Supplements}

\title{Formation of Massive Black Holes in Galactic Nuclei: Runaway Tidal Encounters}
\author[N.~C.~Stone, A.~H.W.~K{\" u}pper and J.~P.~Ostriker]{Nicholas C. Stone$^{1\star}$, Andreas H.W. K{\" u}pper$^1$, and Jeremiah P. Ostriker$^{1, 2}$
\\$^1$Columbia Astrophysics Laboratory, Columbia University, 
New York, NY 10027, USA
\\$^2$Princeton University Observatory, Princeton, NJ 08544, USA
\\$^\star$Einstein Fellow; nstone@phys.columbia.edu}

\begin{document}

\date{\today}
\maketitle

\begin{abstract}

Nuclear star clusters (NSCs) and supermassive black holes (SMBHs) both inhabit galactic nuclei, coexisting in a range of bulge masses, but excluding each other in the largest or smallest galaxies.  We propose that the transformation of NSCs into SMBHs occurs via runaway tidal captures, once NSCs exceed a certain critical central density and velocity dispersion.  The bottleneck in this process, as with all collisional runaways, is growing the first e-fold in black hole mass.  The growth of a stellar mass black hole past this bottleneck occurs as tidally captured stars are consumed in repeated episodes of mass transfer at pericenter.  Tidal captures may turn off as a growth channel once the black hole reaches a mass $\sim 10^{2-3} M_\odot$, but tidal disruption events will continue and appear capable of growing the seed SMBH to larger sizes.  The runaway slows (becomes sub-exponential) once the seed SMBH consumes the core of its host NSC.  While the bulk of the cosmic mass density in SMBHs is ultimately produced (via the Soltan-Paczynski argument) by episodic gaseous accretion in very massive galaxies, the smallest SMBHs have probably grown from strong tidal encounters with NSC stars.  SMBH seeds that grow for a time $t$ entirely through this channel will follow simple power law relations with the velocity dispersion, $\sigma$, of their host galaxy.  In the simplest regime it is  $M_\bullet \sim \sigma^{3/2}\sqrt{M_\star t / G} \sim 10^{6}M_\odot (\sigma / 50~{\rm km~s}^{-1})^{3/2}(t/10^{10}~{\rm yr})^{1/2}$, but the exponents and prefactor can differ slightly depending on the details of loss cone refilling.  Current tidal disruption event rates predicted from this mechanism are consistent with observations.

\end{abstract}

\section{Introduction}

Can star capture account for the growth of the largest supermassive black holes in the Universe?  No.  Could this process account for the growth of abundant smaller supermassive black holes, or intermediate mass seeds?  The answer to this question appears to be yes, as we argue in this paper.

The central few parsecs of most standard ($M_* > 10^9 M_\odot$) galaxies contain either supermassive black holes (SMBHs), nuclear star clusters (NSCs), or both of these components.  In some cases the NSCs are distinct entities and in other cases they appear to be a continuation of the inner spheroidal component.  In either case, NSCs typically represent regions of high stellar density, often with a steep, cusp-like profile.

\begin{figure}
\includegraphics[width=85mm]{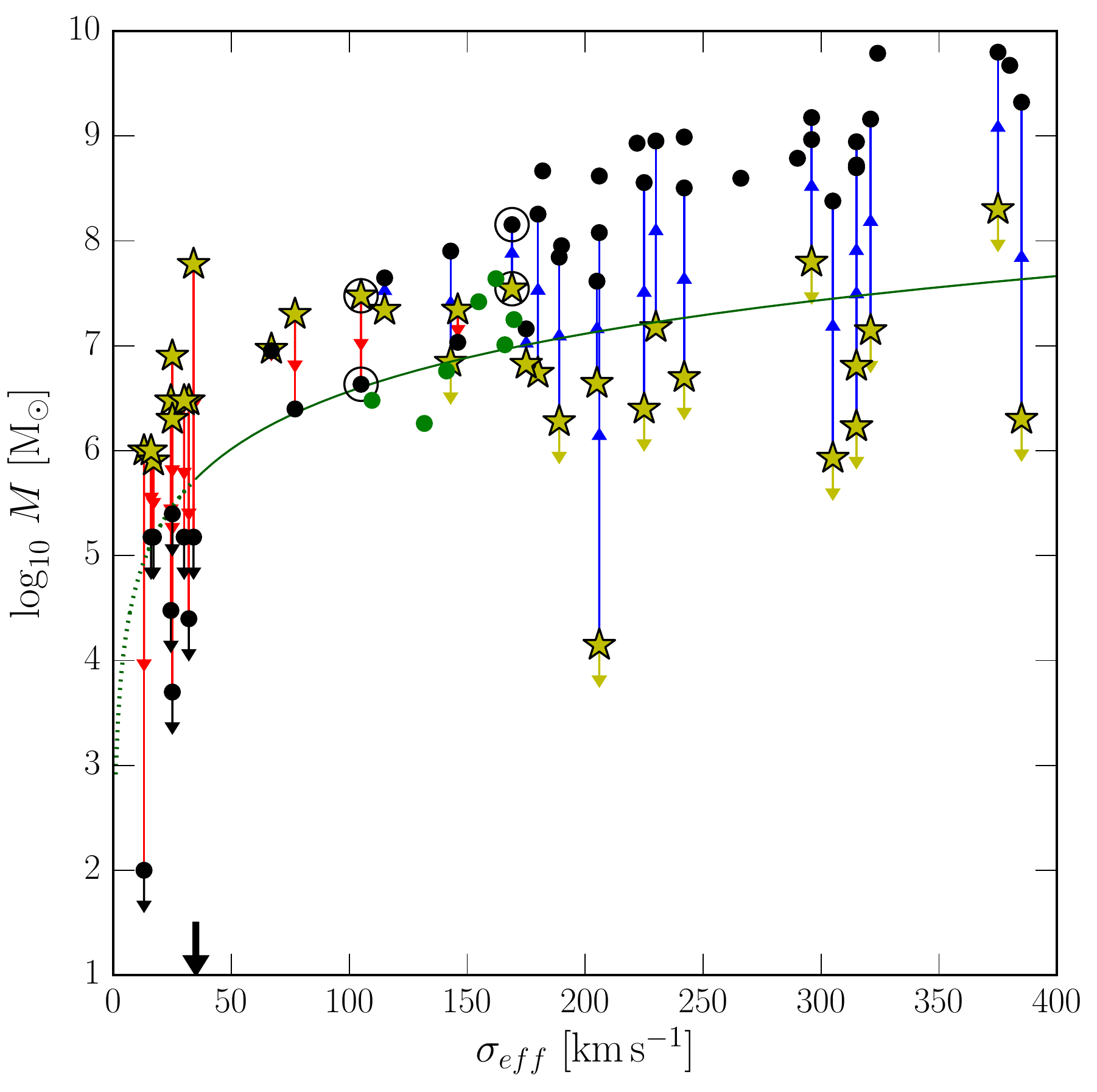}
\caption{Nuclear star cluster (yellow stars) and supermassive black hole (black and green dots) masses in local universe galaxies, plotted against host galaxy velocity dispersion.  A ``break point'' at $\sigma_{\rm eff} \approx 100~{\rm km~s}^{-1}$ is visible, below which NSCs dominate and above which SMBHs dominate.  In low $\sigma_{\rm eff}$ galaxies, NSC masses appear to follow a power law that turns over for $\sigma_{\rm eff} \gtrsim 100~{\rm km}~{\rm s}^{-1}$; this power law is resumed in larger galaxies by the SMBH masses, which fall off sharply for $\sigma_{\rm eff} \lesssim 100~{\rm km~s}^{-1}$.  SMBH dynamical mass measurements and NSC mass data are taken from \citet{GraSpi09, ErwGad12, NeuWal12, KorHo13, GeoBok14}, and two local galaxies of interest (the Milky Way and M31) are marked with circles; these datasets give one SMBH mass (black dots) for every NSC mass.  We also show a different set of SMBH masses estimated with maser disk measurements \citep{Greene+16}; for these galaxies associated NSCs are unconstrained and the SMBH masses are shown with green dots.  The green line shows the ``saturation mass'' predicted for SMBHs formed through the mechanism in this paper, described in more detail in Eq. \ref{eq:MSat}.  Growth of SMBHs above the green line occurs through standard forms of gas accretion \citep{Soltan82}.  The black arrow at $\sigma =35 ~{\rm km~s}^{-1}$ indicates the rough transition below which NSCs are insufficiently dense to produce runaway SBH growth.}
\label{fig:NSCDichotomy2}
\end{figure}

It is well known that, for the more massive galaxies ($M_* > 10^{11} M_\odot$), the mass of the BH component is roughly linearly related \citep{MarHun03, McCMa13, KorHo13} to the mass in the spheroidal stellar component of the galaxy, but lower mass systems can also contain SMBHs with masses below the linear relation, and for very low mass galaxies the SMBHs are apparently absent in general, with notable exceptions \citep{Baldas+15}.  However, for galaxies with masses below this transition ($M_* < 10^{11} M_\odot$), the central regions typically contain NSCs having masses roughly proportional to the spheroidal stellar mass, with values similar to the SMBH mass at the transition point, and with the NSCs being less prominent than the SMBHs above the transition.  NSCs are typically absent in the largest galaxies, and only coexist with SMBHs in galaxies from an intermediate mass range \citep{Georgi+16}.  Fig. \ref{fig:NSCDichotomy2} illustrates this dichotomy with observational data from nearby galactic nuclei, plotted against the velocity dispersion of the host galaxy.

An overly simple explanation for this behavior would postulate that all standard galaxies start out with nuclear star clusters proportional to the spheroidal mass, but that above some critical mass (or escape velocity), dynamical processes transform a larger and larger fraction of the stars in the NSC to massive seed central BHs. That would explain in a simple fashion the continuity observed between centrally located NSCs and central BHs.  The purpose of this paper is to outline the dynamical processes that could make this cartoon scenario a plausible physical process.  We focus in particular on tidal capture and disruption of stars as a runaway growth channel for stellar mass black holes (SBHs).  

The origin of NSCs is not the primary topic of this paper, but the hypothesis that they form from the infall of globular clusters (GCs) due to dynamical friction \citep{Tremai+75, Gnedin+14} is consistent with the dynamical processes that we will be treating in this paper, and is generally capable of growing NSCs to $\sim 10^6 M_\odot$ in less than one billion years\footnote{Though we note that the relationship between NSCs and GCs may be more complex, and some GCs may even descend from tidally stripped NSCs \citep{Boker08}.}.  An alternative formation channel is {\it in situ} growth through star formation \citep{Milosa04, Antoni+15}.  It is not clear which of these two mechanisms dominates NSC formation, and the data is often consistent with contributions from both channels \citep{Leigh+12, Antoni+12}.  However, it is clear that {\it in situ} star formation continues at some level in most nuclear star clusters\footnote{For example, the Milky Way NSC exhibits a large spread in stellar metallicity \citep{Do+15}, and contains two disks of young stars formed $\sim 6~{\rm Myr}$ ago \citep[and references therein]{Genzel+10}.}, and therefore provides a natural source of SBHs in NSCs.  

The primary physical mechanism that we will consider, tidal capture, is well known. One star can capture another into a bound orbit if a close encounter between the two puts more energy into tidal perturbations than the positive relative energy of these two stars at large separation. The tidal capture mechanism was first pointed out by \citet{Fabian+75}, with an early computation by \citet{PreTeu77} and a more refined modal analysis given by \citet{LeeOst86}.  Later work analyzed subsequent orbital evolution both prior \citep{Mardli95a, Mardli95b} and subsequent \citep{Lai96, Lai97} to the onset of internal dissipation.  

If SBHs are retained, or are subsequently formed, in dense stellar systems such as GCs or NSCs, then mass segregation will bring them to the central, densest part of the system, whereupon they will be well situated to tidally capture the much more abundant low mass, normal stars. The dramatic influence that the resulting tidal capture binaries can have on the dynamical evolution of such clusters has been studied extensively \citep{Lee87, Statle+87, LeeOst93, Kim+98}.  But the specific possibility that a runaway of successive tidal captures would lead to the formation of a massive black hole, first proposed by \citet{MilDav12}, has not been quantified in detail. We will provide some first calculations of the circumstances required for such a runaway to occur in this paper, as well as the expected endpoint of the runaway, returning to a more detailed calculation in subsequent work.  The preliminary analytic and numerical treatment presented here indicates that the physical requirements for runaway tidal captures are first satisfied for galactic nuclei near the NSC/SMBH transition at black hole masses of $\sim 10^{5-6} M_\odot$ and stellar velocity dispersions $\sim 35~{\rm km~s}^{-1}$. 

Supermassive black hole (SMBH) seed formation is generally studied in the context of the high redshift universe, and three leading candidate scenarios currently exist \citep[see][for general reviews]{Volont10, Sesana12}: stellar remnant BHs left over following the deaths of Pop III stars, the direct collapse of gas in small halos, and runaway stellar collisions in dense star clusters.  The Pop III scenario has the advantages of concreteness and ubiquity, but has been cast into doubt by recent simulations that find Pop III stars may be much less massive than was previously thought due to fragmentation during their formation \citep{Clark+11, Greif+11}.  Even if simpler estimations for the masses of the first stars ($\sim 10^{2-3} M_\odot$) are correct \citep{Abel+02}, this scenario produces the lowest mass SMBH seeds.  Direct collapse of gas in early halos has the advantage of producing much larger seeds, $\sim 10^{5-6} M_\odot$, but will be strongly suppressed by small amounts of coolants, primarily molecular hydrogen \citep{Visbal+14}.

The mechanism in this paper is a specific example of the third channel for SMBH seed formation: runaway collisions of stellar mass objects in dense stellar environments \citep{Sanders70, BegRee78, Ebisuz+01, Gurkan+04}, such as GCs or NSCs.  Past studies of this channel predicted SMBH seeds of widely varying mass, depending on which variant of the stellar runaway proceeds: $M_\bullet \sim 10^{2-5} M_\odot$.  
We will focus primarily on a relatively unexplored variant of the runaway collision scenario, specifically the process of runaway tidal capture and the resulting growth of a massive object, although we comment briefly on more traditional versions of this scenario as well.  

Our work focuses on tidal capture because that process has the largest cross-section of any of the dynamical mechanisms we consider. It is therefore vital in initiating the runaway growth of black holes that start with roughly $10 M_\odot$ SBH
seeds. But, as we will note later, such processes may not ultimately dominate the total mass growth of our black holes; and consideration of conventional tidal destruction events as initially envisaged by \citet{MagTre99} under the same circumstances would also grow SBHs to comparable sizes provided a runaway can begin.

In \S 2 we discuss the different star capture channels available to grow a SBH in a dense star cluster, and summarize existing observations of NSCs, which appear to provide the most favorable environments for runaway SBH growth.  In \S 3 we analytically explore the collisional growth of SBHs in NSCs.  In \S 4 we use idealized numerical studies to validate our analytic estimates and to test the importance of SBH multiplicity.  In \S 5 we examine the slowing and eventual termination of SBH growth through star capture.  In \S 6 we discuss future observational tests of our model for massive black hole growth, and in \S 7 we offer conclusions.

\section{Physical Processes in Dense Stellar Systems}
\begin{figure*}
\includegraphics[width=170mm]{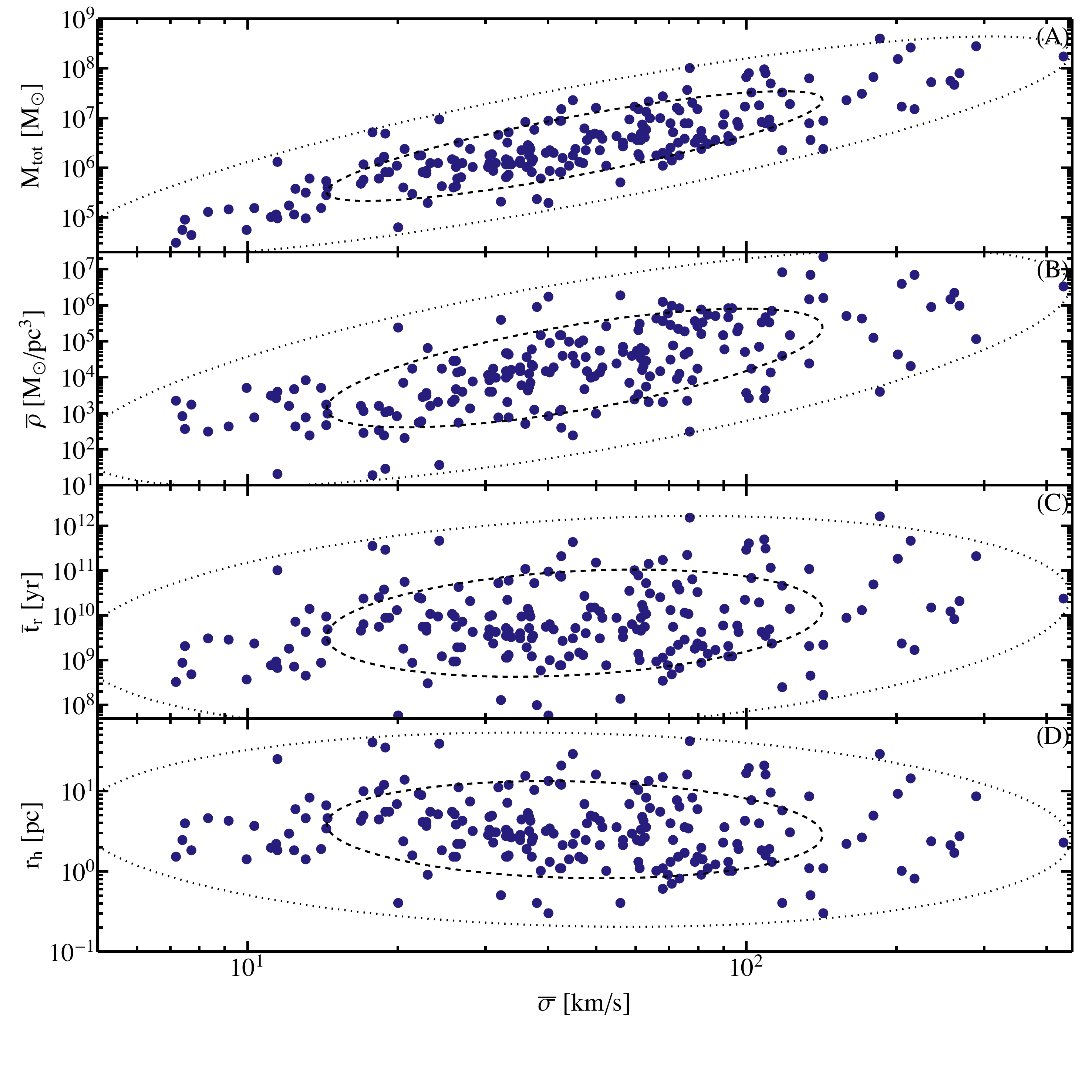}
\caption{Nuclear star cluster quantities of interested plotted against the mean 1D velocity dispersion $\bar{\sigma}$.  In this figure we focus on ``cluster-averaged'' quantities.  Panel (A) shows total cluster mass $M_{\rm tot}$; panel (B) shows mean cluster density $\bar{\rho}$; panel (C) shows mean cluster relaxation time $\bar{t}_{\rm r}$; panel (D) shows half mass radius $r_{\rm h}$.  In all panels the dashed and dotted lines show $1\sigma$ and $2\sigma$ contours from fitted 2D Gaussians.  Data is taken from \citet{Boker+04, Cote+06, GeoBok14}.}
\label{fig:NSCData}
\end{figure*}

To set the stage for our discussion it is useful to summarize the properties of the dense (primarily nuclear) star clusters that we will be considering, and then to outline the primary physical processes occurring in these systems.  These clusters can, with a great degree of simplification, be characterized by three parameters: mass $M_{\rm tot}$, core radius $r_{\rm c}$, and half mass radius $r_{\rm h}$.  We show distributions of some of these quantities in actual NSCs (taken from \citealt{Boker+04, Cote+06, GeoBok14}) in Fig. \ref{fig:NSCData}, along with mean density $\bar{\rho} \equiv 3M_{\rm tot}/(4\pi r_{\rm h}^3)$, mean relaxation time $\bar{t}_{\rm r} \equiv 0.34 \sigma^3/(G^2\bar{M}_\star \bar{\rho} \ln\Lambda)$, and average velocity dispersion\footnote{Throughout this paper, all velocity dispersions used are one-dimensional.} $\bar{\sigma}^2 \equiv GM_{\rm tot}/(3r_{\rm h})$.  Here we take the Coulomb logarithm $\Lambda = 0.4 M_{\rm tot}/\bar{M}_\star$ and assume mean stellar mass $\bar{M}_\star = M_\odot$ for simplicity.   We provide more detail on these data sets, as well as best fit relations and variances between them, in Appendix \ref{app:data}.

\begin{figure*}
\includegraphics[width=170mm]{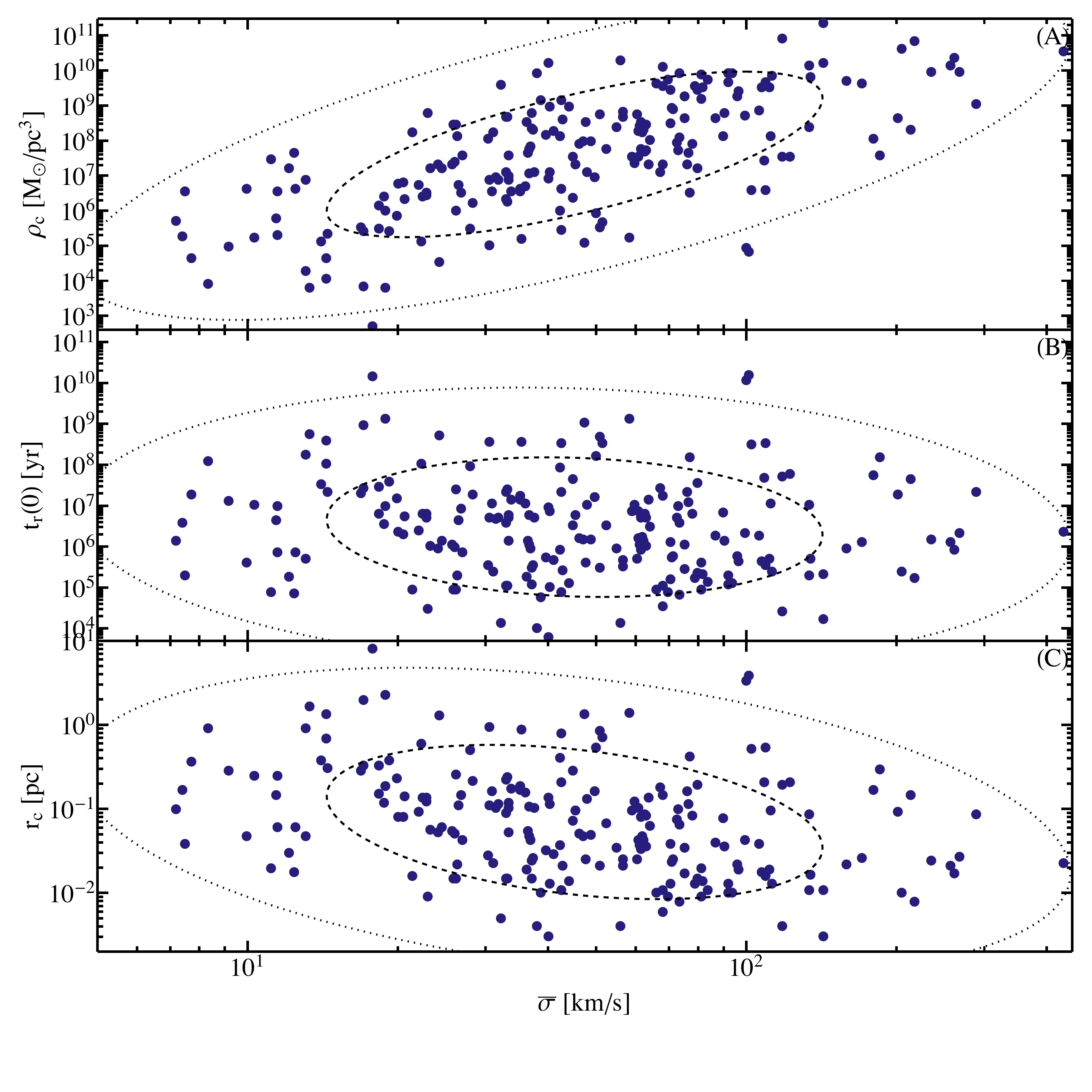}
\caption{The same as Fig. \ref{fig:NSCData} but for three ``cluster-central'' quantities: cluster central density $\rho_{\rm c}$ (panel A), cluster central relaxation time $t_{\rm r}(0)$ (panel B), and core radius $r_{\rm c}$ (panel C).  These three variables are all derived from the concentration parameter $C$, which is not measured with great precision and which in many cases represents a lower limit on the true concentration; for more discussion see Appendix \ref{app:data}.}
\label{fig:NSCData2}
\end{figure*}

Although all of these quantities have significant variance from their means, we see that NSCs typically have $20~{\rm km~s}^{-1} \lesssim \bar{\sigma} \lesssim 150~{\rm km~s}^{-1}$.  Other work has established that this is generally correlated with, and comparable to, the velocity dispersion of the host galaxy \citep{Leigh+15}.  The characteristic sizes of NSCs are typically $r_{\rm h} \sim 1-10 {\rm pc}$, giving a wide range in $\bar{\rho}$.  Relaxation times at the half-mass radius $t_{\rm r}(r_{\rm h})$ are under a Hubble time for $\approx 60\%$ of our sample.  In reality, NSCs are not simple two-parameter systems, and in particular exhibit a wide range of concentration parameters $C \equiv r_{\rm h}/r_{\rm c}$, where $r_{\rm c}$ is the core radius of a roughly isothermal system, and $C \sim 10-100$ \citep{GeoBok14}, implying central densities $\rho_{\rm c} \sim C^2 \bar{\rho}$ that are a few orders of magnitude larger than $\bar{\rho}$.  We plot the distributions of ``cluster-central'' quantities in Fig. \ref{fig:NSCData2}.

We note here that the definition of an NSC is not a completely rigorous one, and whether or not a galaxy's central surface brightness profile contains an NSC can be a subjective question.  The data quality of the innermost observed isophotes, the number of components used in fitting the surface brightness profile, and assumptions about the radial dependence (or lack thereof) of a mass-to-light ratio are all important ingredients in determining whether there is an inner light excess that can be identified as an NSC.  To some extent, however, this debate is a semantic one from the perspective of our calculations, which depend primarily on two measurable quantities: stellar density and stellar velocity dispersion.  Whether or not the center of a galaxy (on $\sim {\rm pc}$ scales) is formally an NSC or is merely an inward continuation of a larger-scale surface brightness power law will not affect our results, so long as it has the same density and velocity dispersion in both scenarios.

If a NSC lacks a massive black hole, how can it acquire one?  In the remainder of this section we consider two different ``collisional runaway'' channels.  The first is the classical prompt collisional runaway that is sometimes theorized to occur in denser star clusters at high redshift; we summarize past literature on this channel and explain why it is unlikely to occur for the NSCs we observe at low redshift.  The second is a relatively unexplored delayed collisional runaway channel, which forms the basis for the remainder of this paper.

\subsection{Prompt Collisional Runaways}

After the formation of a star cluster, its constituents will evolve toward energy equipartition: heavy cluster members, such as massive young stars or SBHs, will lose orbital energy to lighter members, and will mass segregate to the center of the cluster.  Recent N-body simulations indicate that star clusters may never reach a state of full energy equipartition, but nonetheless partial mass segregation does occur \citep{TreMar13}.  If we consider a two component cluster made up of light masses $M_\ell$ and heavy masses $M_{\rm h}$, this occurs on a timescale $t_{\rm seg} \sim (M_\ell / M_{\rm h})t_{\rm r}$, where $t_{\rm r}$ is the energy relaxation time.  

Extremely dense star clusters can undergo a prompt collisional runaway among massive young stars if the mean time between collisions, $t_{\rm coll}$, becomes significantly less than the main sequence lifetime of very massive stars, $\sim 10^6~{\rm yr}$.  If this condition is satisfied, runaway collisions between main sequence stars will form a supermassive star of mass $M_\bullet \sim 10^{-3}M_{\rm tot}$  \citep{Gurkan+04, Goswam+12}, which will eventually collapse to an intermediate mass black hole (IMBH) due to the onset of GR instability \citep[e.g.][]{ShaTeu83}.  
The approximate central number densities required for formation of a supermassive star in this type of prompt collisional runaway are $n_{\rm c} \gtrsim 10^9 M_\odot~{\rm pc}^{-3}$, roughly two orders of magnitude greater than the highest seen in Fig. \ref{fig:NSCData}.  Clusters could form at this density at high redshift due to dissipation in gas flows \citep{Katz+15}, or alternatively could reach central densities of this order via Spitzer instability \citep{Spitze69} or a broader core collapse.  

However, in the low redshift universe, this process is derailed by the nonzero metallicity of stellar participants.  Collision products will shed most of their mass through line-driven winds, thus forming stellar-mass compact remnants in supernova explosions; even very low-metallicity progenitors of modest size ($Z=0.001Z_\odot$; $M \sim 500 M_\odot$) lose enough mass via winds to prevent IMBH formation \citep{Glebbe+09}.  



\subsection{Delayed Collisional Runaways}
Let us assume that there is no prompt collisional runaway; instead, massive stars segregate to the center of the star cluster and die in supernovae there, leaving behind a smaller population of compact object remnants (most neutron stars and likely some SBHs will escape in natal kicks).  These SBHs will swap into primordial binaries in binary-single encounters, and the larger interaction cross-sections of these binaries will eventually lead to ejections of many black holes.  As this process repeats, the number of SBHs declines until either $0-1$ remain, or the SBH-SBH interaction time has become very long (this competition between SBHs is analyzed through N-body integrations in \S \ref{sec:numerics}).  Even if our NSC is unlucky and loses all its SBHs in this process, the growth of NSCs over time (either through {\it in situ} or cluster accretion processes) implies that it will eventually regain one.

Several different channels exist for the growth of an SBH in a gas-free star cluster, which we describe in quantitative detail in the appendices of this paper.  The four growth channels we consider are the following, listed in order of increasing cross-section and probability:
\begin{enumerate}
\item Gravitational wave capture of compact objects (Appendix \ref{sec:GW}): two compact objects on hyperbolic orbits can capture into a bound orbit through emission of GWs.  
\item Direct collision with main sequence stars: a SBH can directly collide with a main sequence star if the pericenter of their encounter, $R_{\rm p}$, is less than the stellar radius $R_\star$.  The end product of such a collision is highly uncertain.
\item Tidal disruption of main sequence stars (Appendix \ref{sec:TDE}): a SBH will tidally disrupt a main sequence star of mass $M_\star$ if $R_{\rm p} < R_{\rm t} \equiv R_\star(M_\bullet / M_\star)^{1/3}$, where we have defined the tidal radius $R_{\rm t}$.  Up to half the stellar mass can accrete onto the SBH.
\item Tidal capture of main sequence stars (Appendix \ref{sec:TC}): a SBH will tidally capture an unbound star into a bound orbit (by dumping excess orbital energy into tidally-excited oscillation modes) if $R_{\rm p} < \lambda R_{\rm t}$, where $\lambda \approx 2$ is a function of stellar structure and cluster $\sigma$.  Depending on subsequent orbital evolution of this tidal capture (TC) binary, the star may be consumed by the SBH.
\end{enumerate}
Collisional cross-sections are often enhanced by gravitational focusing of orbits, so that the impact parameter $b$ for a close-approach distance of $R_{\rm p}$ is given by
\begin{equation}
b = R_{\rm p} \sqrt{1+\frac{2G(M_1+M_2)}{R_{\rm p}\sigma^2}},
\end{equation}
where $M_1$ and $M_2$ are the masses of the impactors.  Generic encounter rates can be calculated at leading order as $\dot{N}=n\Sigma v$, where $n$ is the density of targets, $v$ is the relative velocity between SBHs and their targets (typically $v\approx \sigma$), and $\Sigma = \pi b^2$.

Of these processes, GW capture of compact objects has the smallest cross-section by far, and will generally be disregarded in this paper as a growth mechanism for SBHs (see Appendix \ref{sec:GW} for a more detailed discussion).  In the gravitationally focused limit, the rate of growth due to direct collisions is at most $\dot{M}_\bullet \propto M_\bullet$, but we neglect this channel both because it is unclear whether significant stellar mass will accrete, and because the exponential growth that is theoretically possible from direct collisions is slower than the super-exponential growth allowed by tidal processes.  Both tidal capture and tidal disruption scale as $\dot{M}_\bullet \propto M_\bullet^{4/3}$, a curve of super-exponential growth that formally goes to infinite mass in finite time.  Because the tidal capture process has the largest cross-section, tidal capture runaways are the most promising process for growing an IMBH in NSC-like clusters that lack one\footnote{Although under certain circumstances tidal disruptions may dominate; see \S \ref{sec:outcomes}.}, and we explore the onset and development of these runaways in the next section. 

\section{Tidal Capture Processes in Realistic Nuclear Star Clusters}
\label{sec:clusterObs}
In this section, we summarize our theoretical models for realistic NSCs, and then apply simple analytic models for collisional growth of SBHs in these environments.  Although in principle tidal capture (or other types of) runaways can happen in open or globular clusters as well, runaway SBH growth is ultimately density-limited, and we therefore focus on the densest observed stellar systems in the universe: NSCs.  A worthwhile caveat is that more dense star clusters may have existed at high redshift, with central subclusters driven to even higher densities through gas dissipational processes \citep{Davies+11, Leigh+14}.

\subsection{Cluster Structure}
\label{sec:clusterStruct}

We use the three parameter potential-density pair described in \citet{StoOst15} to approximate the structure of all of these star clusters.  Specifically, the stellar density profile of a cluster with central density $\rho_{\rm c}$, core radius $r_{\rm c}$, and halo radius $r_{\rm h}$ (equivalent to the half-mass radius when $r_{\rm c} \ll r_{\rm h}$) is
\begin{equation}
\rho(r) = \frac{\rho_{\rm c}}{(1+r^2/r_{\rm c}^2)(1+r^2/r_{\rm h}^2)},
\end{equation}
and the gravitational potential 
\begin{align}
\Phi = &-\frac{2GM_{\rm tot}}{\pi (r_{\rm h}-r_{\rm c})} \Bigg[\frac{r_{\rm h}}{r}\arctan\left(\frac{r}{r_{\rm h}}\right) - \frac{r_{\rm c}}{r}\arctan \left( \frac{r}{r_{\rm c}} \right)\notag \\
& + \frac{1}{2}\ln \left(\frac{r^2+r_{\rm h}^2}{r^2+r_{\rm c}^2} \right) \Bigg],
\end{align}
where the total cluster mass $M_{\rm tot} = 2\pi^2r_{\rm c}^2r_{\rm h}^2\rho_{\rm c}/(r_{\rm h}+r_{\rm c})$.  This potential-density pair is designed as an analytically tractable alternative to single-mass King models \citep{King66}; it is appropriate for use here because observed NSCs are reasonably well fit by King models, with varying degrees of concentration $C \equiv r_{\rm h} / r_{\rm c}$.  \citet{GeoBok14} find that over half ($58\%$) of their large NSC sample is best fit by highly concentrated King models, with $C \sim 100$.  

When $r_{\rm c} \ll r_{\rm h}$, the central 1D velocity dispersion is
\begin{equation}
\sigma_{\rm c}^2 = \frac{2GM_{\rm tot}(\pi^2/8-1)}{\pi r_{\rm h}},
\end{equation}
an approximation we shall use for the remainder of this paper (when $r_{\rm c} \sim r_{\rm h}$, the velocity dispersion is slightly higher than this value).  Finally, the central relaxation time is
\begin{equation}
t_{\rm r}(0) = \frac{0.39}{\ln \Lambda} \frac{M_{\rm tot}}{M_\star} \frac{\sqrt{r_{\rm c} r_{\rm h}}}{r_{\rm h}+r_{\rm c}}\sqrt{ \frac{r_{\rm c}^3}{GM_{\rm tot}} }. \label{eq:trelax}
\end{equation}
At distances $r \ll r_{\rm h}$, $t_{\rm r}(r) \approx t_{\rm r}(0)(1+r^2/r_{\rm c}^2)$.

The plausibility of collisional runaways in star clusters of fixed mass $M_{\rm tot}$ and halo (roughly speaking, half mass) radius $r_{\rm h}$ depends critically on the size of the core, as $\rho_{\rm c} \propto r_{\rm c}^{-2}$.  Clusters lacking central heat sources will, over $\approx 300$ central relaxation times \citep[under the assumption of equal-mass stars in the cluster]{Cohn80}, enter a state of core collapse, as heat is slowly conducted out of the core, and $r_{\rm c}$ approaches zero.  In practical terms, the presence of primordial binaries in the core can arrest this collapse (scattering of single stars off a hard binary will harden the binary further and heat the core, offsetting conductive heat losses).  \citet{MilDav12} have argued that star clusters that possess $\sigma \gtrsim 40~{\rm km~s}^{-1}$ can avoid this problem by ``burning through'' their entire population of primordial binaries.  

To better quantify this argument (and the possible existence of a $\sigma \sim 40~{\rm km~s}^{-1}$ threshold for strong susceptibility to core collapse), we consider an equal mass binary with total mass $M_{\rm b} = 2M_\star$ in the core of a cluster with mean stellar mass $\bar{M}_\star$.  If this binary has a binding energy greater in magnitude than $\bar{M}_\star \sigma^2$, the outcome of many cumulative three-body interactions will be to statistically harden it \citep{Heggie75}, heating the cluster and supporting it against core collapse.  The rate at which the binary loses its binding energy $E_{\rm b}$ to the cluster core is \citep{Spitze87}
\begin{equation}
\frac{{\rm d}E_{\rm b}}{{\rm d}t} \approx 1.3 n_{\rm c} G^2 M_\star \bar{M}_\star (2M_\star + \bar{M}_\star)/\sigma.
\end{equation}
Here the cluster central number density $n_{\rm c} \equiv \rho_{\rm c}/\bar{M}_\star$.  Notably, this hardening rate is independent of binary semimajor axis $a_{\rm b}$.  Slight gradients in the cluster temperature profile will conduct excess heat from the core to the outskirts at a rate $\dot{E}_{\rm cond} = \frac{1}{2}A_{\rm cond}M_{\rm c}\sigma_{\rm c}^2/t_{\rm r}(0)$, so if the cluster core contains $N_{\rm b}$ binaries at any point in time, it will adjust to a steady-state size where heating balances conductive cooling and 
\begin{align}
r_{\rm c} \approx & 2.5 r_{\rm h} \frac{N_{\rm b}}{A_{\rm cond}\ln \Lambda} \frac{M_\star(2M_\star + \bar{M}_\star)}{M_{\rm tot} \bar{M}_\star} \\
\approx& 0.4\frac{N_{\rm b}}{A_{\rm cond}\ln \Lambda} \frac{GM_\star(2M_\star + \bar{M}_\star)}{\sigma_{\rm c}^2 \bar{M}_\star} .\notag \label{eq:rcEq}
\end{align}
The dimensionless conductivity constant $A_{\rm cond} \approx 1.5 \times 10^{-3}$ for single-mass clusters \citep{Cohn80}.  Using this approximation and taking $\ln\Lambda = 12$, we find a central cluster density in steady state of 
\begin{align}
n_{\rm c} \approx & 3 \times 10^{11}~{\rm pc}^{-3} N_{\rm b}^{-2} \left(\frac{r_{\rm h}}{\rm pc}\right)^{-3} \left(\frac{M_{\rm tot}}{10^6 M_\odot} \right)^3 \left(\frac{\bar{M}_\star}{M_\odot} \right) \\ \notag
& \times \left(\frac{M_\star}{M_\odot} \right)^{-4}.
\end{align}
This formula represents an upper limit on the central density of any collisional stellar system, and it decreases rapidly with the mass of central binaries (note, however, that it assumes an equal-mass binary). 

The burn time for an individual binary to reach its minimum separation $a_{\rm min}$ in such a cluster core is 
\begin{align}
t_{\rm burn} \approx& 4.2 \sqrt{\frac{r_{\rm h}^2}{GM_{\rm tot}}} \frac{r_{\rm h}}{a_{\rm min}} \left(\frac{N_{\rm b}}{A_{\rm cond}\ln\Lambda} \right)^2 \left(\frac{M_\star}{\bar{M}_\star} \right)^2 \\
& \times \frac{M_\star}{M_{\rm tot}} \frac{2M_\star+\bar{M}_\star}{M_{\rm tot}}. \notag
\end{align}
Assuming a fraction $f_{\rm b}$ of all cluster stars are born into primordial binaries, we find a total burn time of
\begin{equation}
T_{\rm burn} \approx 8 \times 10^8 ~{\rm yr}~ N_{\rm b} \left(\frac{a_{\rm min}}{R_\odot} \right)^{-1} \left(\frac{r_{\rm h}}{\rm pc} \right)^{5/2} \left(\frac{M_{\rm tot}}{10^6 M_\odot} \right)^{-3/2} \frac{M_\star}{M_\odot} \frac{f_{\rm b}}{0.07}
\end{equation}
for a cluster to burn every one of its primordial binaries; in the last equation we set $M_\star = \bar{M}_\star$ for simplicity.  We note that our choice of $a_{\rm min} = R_\odot$ is likely quite conservative, as primordial binaries will tend to eject in strong three-body encounters once they reach a separation where their orbital speed is comparable to $\sigma$.  Therefore we expect most NSCs to be able to burn through their primordial binaries in a Hubble time, as argued by \citet{MilDav12}, provided they can reach the core on shorter timescales.

In reality, however, binaries are continuously transported to the core by the slow process of dynamical friction in an extended NSC.  Assuming that the orbital speed for binaries is at most mildly supersonic, we can write an approximate dynamical friction timescale for a binary on a circular orbit:
\begin{align}
T_{\rm DF} = &\frac{3\sigma^3}{4\sqrt{2\pi}G^2\bar{\rho} M_\star \ln \Lambda}\\
\approx & 1.6 \times 10^{10}~{\rm yr}~ \sigma_{40}^3 m_{\rm b, 2}^{-1} \bar{\rho}_4^{-1}, \notag
\end{align}
where we use the 1D velocity dispersion, take $\ln \Lambda \approx 3.1$, and define an average cluster density $\bar{\rho} = \bar{\rho}_4 \times 10^4~M_\odot/{\rm pc}^3$.  We therefore conclude that only relatively dense star clusters, with $\bar{\rho} \gtrsim 10^5~M_\odot/{\rm pc}^3$, have both sufficient energy to burn through their primordial binaries and the ability to do this in a Hubble time.

Another complication to the simple cluster model presented above is the possible presence of a ``cavity'' in the center of the cluster, excavated by the scattering interactions of main sequence stars with a dense subcluster of compact objects.  We can estimate the size of this cavity by calculating how long it takes for a star to have a sufficiently close (focused) encounter with a SBH such that it receives a kick velocity greater than the core escape velocity
\begin{equation}
v_{\rm esc}(0) = \frac{2}{\sqrt{\pi}} \sqrt{\frac{GM_{\rm tot}}{r_{\rm h}-r_{\rm c}} } \sqrt{\ln (r_{\rm h}/r_{\rm c})}.
\end{equation}
Here we assume that we are looking at older populations of Spitzer-stable SBHs; this analysis does not apply at very early times in the cluster's history when a Spitzer-unstable SBH subcluster may exist.

\begin{figure}
\includegraphics[width=85mm]{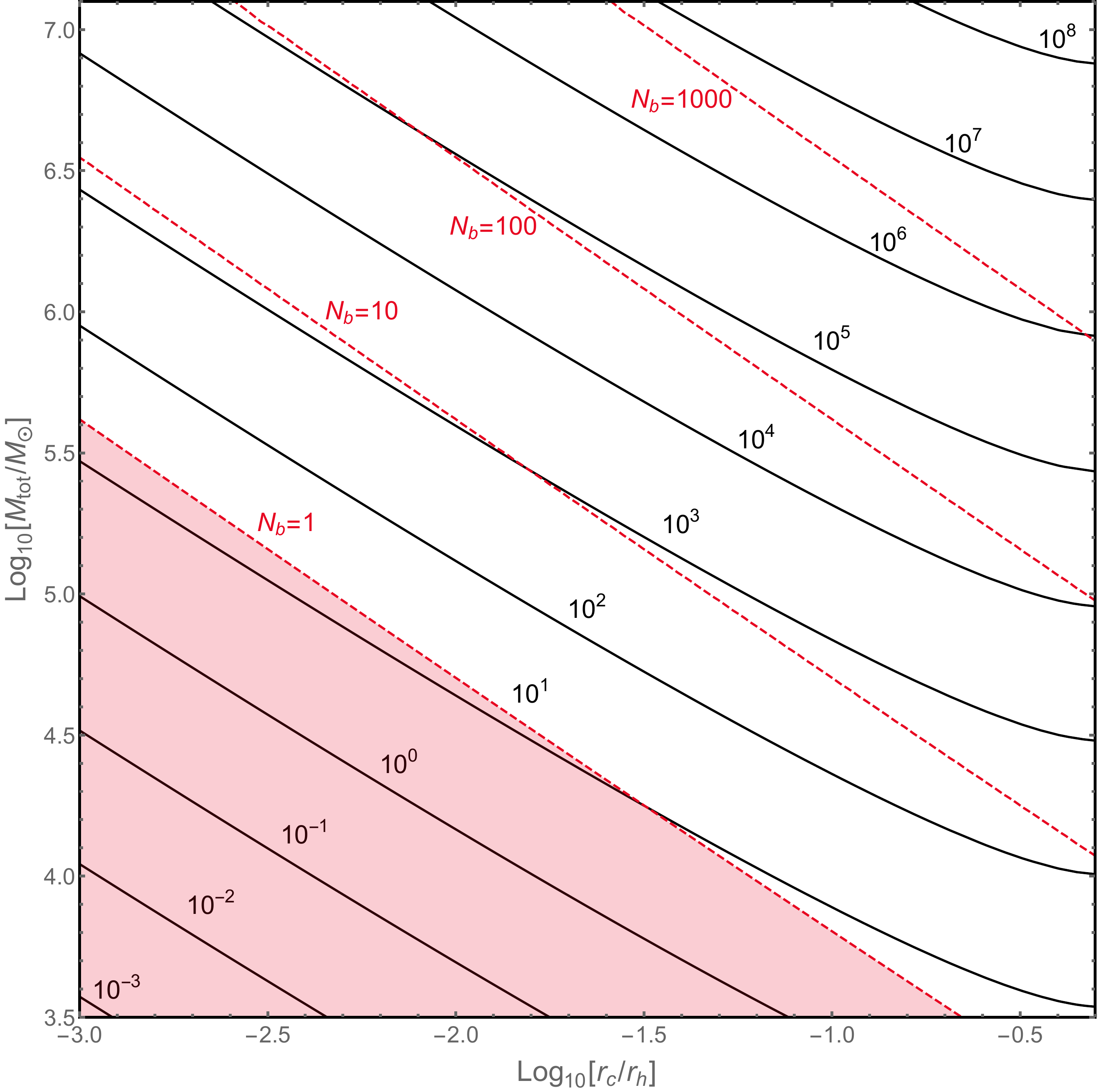}
\caption{A contour plot of the ratio of the stellar ejection time (due to SBH scatterings) to the central relaxation time.  Contours of $\log_{10}(t_{\rm ej}/t_{\rm r}(0))$ are labeled and shown as black solid lines.  Here we assume that there are $N_\bullet = 100$ SBHs all of mass $M_\bullet 10M_\bullet$; even with this unrealistically large late-time SBH population, generally $t_{\rm ej}/t_{\rm r}(0) \gg 1$ (i.e. a central cavity in the stellar profile does not form).  We also plot contours of equilibrium $r_{\rm c}/r_{\rm h}$ for a cluster of a given mass heated by $N_{\rm b}$ binaries of mass $M_{\rm b}=2M_\odot$ (Eq. \ref{eq:rcEq}).  These are shown as dashed red lines with associated labels, and the shaded red region is a portion of parameter space forbidden provided there is at least one binary in the cluster core.  Cavity formation is disfavored.}
\label{fig:cavity}
\end{figure}

By assumption, the $N_\bullet$ SBHs are in energy equipartition, with $\sigma_\bullet^2 = \sigma_{\rm c}^2M_\star/M_\bullet$; this implies that they inhabit a characteristic radius $r_\bullet = r_{\rm c}\sqrt{9(\pi^2/8-1)(M_\star / M_\bullet)/2}$.  A star-SBH scattering event will likely result in the star being ejected from the cluster core if the closest approach is within $r_{\rm ej} = (\pi / 2)(M_\bullet / M_{\rm tot})r_{\rm h}/\ln(r_{\rm h}/r_{\rm c})$.  By taking the gravitationally focused limit and the number density of SBHs, we can compute a per-orbit probability for a star to be ejected, $P_{\rm ej}$.  The ejection timescale $t_{\rm ej} = t_{\rm orb}/P_{\rm ej}$ is, for stars within the cluster core,
\begin{align}
t_{\rm ej} =& \frac{243(\pi^2/8-1)^3}{2^{1/2}\pi^{5/2}}\sqrt{\frac{r_{\rm c}^3}{GM_{\rm tot}}}\left( \frac{r_{\rm c}}{r_{\rm h}} \right)^{5/2} \\
& \times \left( \frac{M_{\rm tot}}{M_\bullet} \right)^3 \left(\frac{M_\star}{M_\bullet} \right)^{3/2} \frac{\ln^{3/2}(r_{\rm h}/r_{\rm c})}{N_\bullet} \notag
\end{align}
In Fig. \ref{fig:cavity}, we make a contour plot of the ratio of $t_{\rm ej}/t_{\rm r}(0)$.  When this ratio is less than unity, a cavity can be evacuated in the central regions of the cluster core, but when it is greater than unity, stars will diffuse inward to refill the core faster than they are ejected.  Interestingly, this ratio does not depend on $r_{\rm c}$ or $r_{\rm h}$ independently, but only on $r_{\rm c}/r_{\rm h}$.  In general, we find that cavities only exist for small and highly concentrated clusters.  However, such clusters cannot exist in a steady state so long as at least one binary exists as a heat source in the cluster core, as is shown in Fig. \ref{fig:cavity}.  Furthermore, it is actually large and highly concentrated clusters that are most relevant for tidal capture runaways, so we neglect the effect of a cavity for the following discussion (but see \S \ref{sec:outcomes2} for how a different type of cavity can form at late times).

In conclusion, we expect high central densities to be achievable through core collapse in many large NSCs ($\sigma \gtrsim 40~{\rm km~s}^{-1}$), although binary burning may persist for greater than a Hubble time unless $\bar{\rho} \gtrsim 10^5~M_\odot~{\rm pc}^{-3}$.  To achieve the high central densities required for runaway SBH growth, we therefore require high $\sigma$ and high (mean) density clusters, although our requirements are satisfied by a large portion of the parameter space in Fig. \ref{fig:NSCData}.  In the following subsection, we quantify better the central density threshold required for the onset of runaway growth.

\subsection{Relative Rates}
\label{sec:relativeRates}

Of the processes described in \S 2 to grow stars or compact objects (gravitational wave capture, direct collision, tidal disruption, and tidal capture), tidal capture and tidal disruption are in general the most promising.  Growth of BHs through GW capture is only possible in extremely deep potential wells where the loss of merger products due to GW recoil can be mitigated \citep{Davies+11}.  The large cross-sections for tidal processes give super-exponential growth ($\dot{M}_\bullet \propto M_\bullet^{4/3}$) more conducive to a runaway than the merely exponential growth provided by direct collisions ($\dot{M}_\bullet \propto M_\bullet$), and without the sizable uncertainty involved in the outcome of a direct collision.  

We therefore examine the rates of tidal capture and tidal disruption relative to other processes that could sabotage the growth of an SMBH seed.  These processes themselves have relative rates given by $\dot{N}_{\rm TC} = \lambda \dot{N}_{\rm TD}$; in Appendix \ref{sec:TC} we argue that $\lambda \approx 2$, but with a weak $\sigma$-dependence.  For the remainder of this subsection we focus on the rate of tidal captures, which we shorthand as $\dot{N}_{\bullet \star}$ for notational clarity, but removal of factors of $\lambda$ enable easy comparison to tidal disruption rates.  We also assume here that a tidally captured star is consumed whole by the SBH; this assumption is explored in much greater detail in \S \ref{sec:outcomes}.

Initially SBHs can grow at a linear rate through tidal captures, reaching a super-exponential pace after they have roughly doubled in size.  If we assume that typical SBHs have $M_\bullet \approx 20M_\odot$ and mean stellar masses in an evolved cluster are $\bar{M}_\star \approx 0.3 M_\odot$, we require $\approx 60$ tidal capture events to reach runaway growth, which corresponds to a tidal capture rate $\dot{N}_{\rm TC} \approx 2\times 10^{-8}~{\rm yr}^{-1}$ if we wish to reach the runaway regime in a Hubble time $t_{\rm H}$.  

Two processes competing with tidal capture for the affection of solitary black holes are binary-single interactions and GW capture.  The former is most relevant in low-$\sigma$ clusters, and the latter in the (possibly short-lived) SBH subclusters produced by mass segregation.  In the gravitationally focused limit relevant for us, the rate of tidal captures by a single SBH is 
\begin{equation}
\dot{N}_{\bullet \star} \approx 2\times 10^{-9}~{\rm yr}^{-1} M_{20}^{4/3} m_\star^{-1/3}r_\star n_{\rm c, 6} \lambda_2 \sigma_{40}^{-1}. \label{eq:TCRate}
\end{equation}
Here we have used for convenience the normalized variables $\sigma_{40}\times 40~{\rm km~s}^{-1} \equiv \sigma$, $n_{\rm c, 6} \times 10^6~{\rm pc}^{-3} \equiv n_{\rm c}$, $M_{20} \times 20M_\odot \equiv M_\bullet$, $m_\star \times M_\odot \equiv \bar{M}_\star$, $r_\star \times R_\odot \equiv \bar{R}_\star$, and $\lambda_2 \times 2 \equiv \lambda$.

Single SBHs will also interact with primordial binaries of semimajor axis $a_{\rm b}$, which can result in capture into the binary or the ejection of some members of this process from the cluster.  We assume these strong interactions occur when pericenter $R_{\rm p} < a_{\rm b}$, and that the logarithmically flat field distribution \citep{Poveda+07} of primordial binary semimajor axes $P(a_{\rm b}) \propto 1/a_{\rm b}$ for $a_{\rm min} < a_{\rm b} < a_{\rm field}$ is truncated within the cluster so that $a_{\rm min} < a_{\rm b} < a_{\rm max}= GM_{\rm b}/(3\sigma^2)$.  Integrating the BH-binary interaction rate over all $a_{\rm b}$ then gives the total BH-binary interaction rate, $\dot{N}_{\bullet \rm b}$, which can be expressed fractionally as 
\begin{align}
\frac{\dot{N}_{\bullet \star}}{\dot{N}_{\bullet \rm b}} = & \frac{\lambda R_\star q^{1/3} \sigma^2}{GM_{\rm b}} \ln \left( \frac{a_{\rm field}}{a_{\rm min}} \right) \label{eq:binaryRate} \\
\approx & 0.37~M_{20}^{1/3}m_\star^{-1/3}r_\star \lambda_2 \sigma_{40}^2 m_{\rm b}. \notag
\end{align}
In the approximate equality above, we have assumed $a_{\rm field} =10^5~{\rm AU}$, $a_{\rm min}=0.01~{\rm AU}$, and have defined $m_{\rm b}\times 2M_\odot \equiv M_{\rm b}$.

\begin{figure}
\includegraphics[width=85mm]{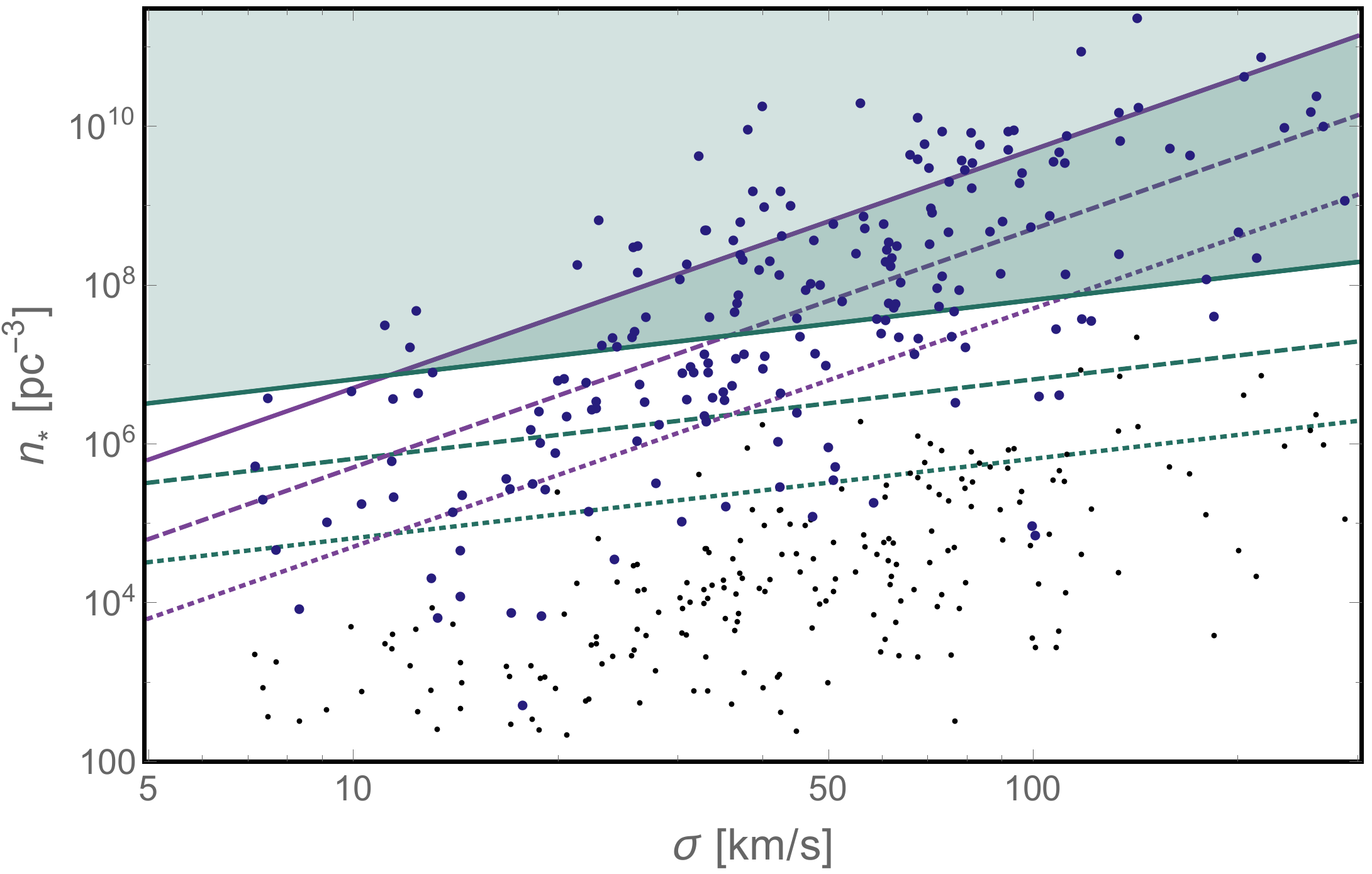}
\caption{Rates of BH-single and BH-binary interactions, plotted over the parameter space of cluster structural parameters $\sigma$ (cluster velocity dispersion) and $n_\star$ (stellar density).  The green lines ($n_\star \propto \sigma^3$) show curves of constant tidal capture rate $\dot{N}_{\bullet \star}$, while the purple lines ($n_\star \propto \sigma$) show curves of constant binary interaction rate $\dot{N}_{\bullet \rm b}$.   Solid lines show rates $10^{-8}~{\rm yr}^{-1}$, dashed lines show rates $10^{-9}~{\rm yr}^{-1}$, and dotted lines show rates $10^{-10}~{\rm yr}^{-1}$.  In all curves we have assumed interactions with solar-type main sequence stars and SBH masses $M_\bullet = 10M_\odot$.  The shaded parameter space above the solid green curve is conducive to runaway growth through tidal capture, although areas above this curve and also above the solid purple curve may see this growth inhibited by more frequent binary interactions.  Binary interactions are less important in the darkly shaded parameter space.  Data points show the \citet{GeoBok14} sample of NSCs; the large blue dots indicate best estimates for central cluster density, while the small black dots illustrate the more robust stellar density at the half mass radius.}
\label{fig:relativeRates}
\end{figure}

We plot the relative rates of these two processes across the parameter space of star clusters in Fig. \ref{fig:relativeRates}, by combining Eqs. \eqref{eq:TCRate} and \eqref{eq:binaryRate} with expressions for the central density ($n_{\rm c}$) and velocity dispersion ($\sigma_{\rm c}$) of realistic clusters.  We see that in order to achieve the conditions necessary for a tidal capture runaway (in less than a Hubble time), it is necessary that the cluster halo (or half-mass) radius $r_{\rm h}$ be very small, typically $\lesssim 1~{\rm pc}$.  Even when this criterion is satisfied, it is likely that binary interactions will occur with greater frequency than tidal captures, unless the cluster is also quite massive $\gtrsim 10^7 M_\odot$. 

Binary-single encounters can prevent a tidal capture runaway from ever beginning, by ejecting all SBHs from the cluster (or at least, the cluster core).  If we imagine an encounter between a main sequence star and a SBH-star binary with semimajor axis $a_{\rm b}$, one of the stars will typically be ejected at a speed $v_{\rm e} \sim \sqrt{GM_\bullet / a_{\rm b}}$, and momentum conservation dictates that the surviving SBH-star system recoils with $v_{\rm ej} \sim (M_\star / M_\bullet) v_{\rm e}$.  This process clearly sets a lower limit on the cluster $\sigma$ capable of retaining SBHs, as the central cluster escape velocity is
\begin{equation}
v_{\rm esc}(0) = \frac{2}{\sqrt{\pi}} \sqrt{\frac{GM_{\rm tot}}{r_{\rm h}-r_{\rm c}} } \sqrt{\ln (r_{\rm h}/r_{\rm c})}.
\end{equation}
The largest binary ejection velocities come from the hardest possible SBH-star binaries, i.e.
\begin{equation}
v_{\rm ej}^{\rm max}= \left(\frac{m_\star}{m_\bullet} \right)^{2/3} \sqrt{\frac{Gm_\star}{R_\star} }.
\end{equation}
If $v_{\rm ej}^{\rm max} \gtrsim v_{\rm esc}(0)$, then it becomes impossible to retain SBHs in the cluster and tidal capture runaways are prevented.  


If we take an angle-averaged distribution of ejection speeds $v_{\rm ej}$ \citep{ValKar06}, we find that encounters between SBHs and primordial binaries eject the newly bound SBH-star pair with speed $\ge v$ at a rate
\begin{equation}
\dot{N}_{\rm ej}(v) = \frac{\pi f_{\rm b}G^2M_{\rm b}^2 n_{\rm c}}{32\sigma v^2 \ln(a_{\rm max}/a_{\rm min})} \left(\frac{M_{\rm b}}{M_\bullet} \right)^{2/3}.
\end{equation}
Clearly, frequent ejections with $v \gtrsim v_{\rm esc}(0)$ will endanger any runaway, but even smaller values of $v$ may abort runaway growth, as SBHs ejected to the low-density cluster halo will not grow through star capture and may take very long times to sink back to the center \citep{Morsche+15}.  We therefore compute a ``core escape velocity'' $v_{\rm c}^2 \equiv 2\Phi(r_{\rm h}) - 2\Phi(0)$, and we find that $\dot{N}_{\rm ej}(v_{\rm c}) > \dot{N}_{\bullet \star}$ when
\begin{align}
\sigma^2 \lesssim& \frac{f_{\rm b}(\pi^2/8-1)}{32\lambda} \frac{1}{\ln(a_{\rm max}/a_{\rm min})(\ln(r_{\rm h}/2^{1/2}r_{\rm c})-\pi/4)} \notag \\
& \times \frac{\bar{M}_\star}{M_\bullet} \frac{G\bar{M}_\star}{\bar{R}_\star} \\
\sim& (1~{\rm km~s}^{-1})^2 \lambda_2^{-1}M_{20}^{-1}m_\star^2 r_\star^{-1}.\notag
\end{align}
In the above, we have assumed $r_{\rm h} \gg r_{\rm c}$; we see that primordial binaries are unlikely to efficiently eject SBHs from NSC cores, though they will be much more effective at this in open and globular clusters.

Even though $\sigma$ and $M_\bullet$ are usually large enough for the SBH to survive ejection in binary-single encounters, these interactions can still endanger a tidal capture runaway because tides capture stars onto highly eccentric orbits.  If the SBH is part of a relatively wide (but still hard, by cluster standards) binary system at the time of capture, then chaotic three-body interactions can ensue, and it is likely that the captured star will quickly scatter to a larger pericenter where it no longer has the chance to undergo strong tidal interactions or circularize, as we discuss further in the next section.  In a conservative sense, therefore, it is best to treat only the darkly shaded part of Fig. \ref{fig:relativeRates} as the part of parameter space where runaway tidal growth is favored.    

\subsection{Outcomes of Star Capture}
\label{sec:outcomes}

After a star has been tidally captured by a BH, it will return to pericenter and will suffer further strong tidal encounters.  However, unlike on the first (unbound) passage, when tidal forces excite oscillatory modes in a previously quiescent star, repeated pericenter passages on a bound orbit can both excite and de-excite oscillation modes.  In the absence of dissipation, the exchange of energy between modes and orbit will be either quasi-periodic and bounded (if the pericenter is large relative to a ``chaos boundary'' produced by the overlap of mode-orbit resonances - see \citealt{Mardli95a}), or will undergo an unbounded random walk (if the pericenter is small relative to the chaos boundary).  In practice, tidal capture in the high-$\sigma$ environment of a NSC requires a small pericenter, and so the mode amplitude will follow a chaotic random walk in the absence of dissipation, often reaching nonlinear sizes \citep{Mardli95b}.  

The first passage between the SBH and an unbound star transfers the following orbital energy
\begin{equation}
\Delta E_0 = \frac{GM_*^2}{R_*} \left( \frac{M_\bullet}{M_*} \right)^2 \sum_{\ell=2, 3, ...}^{\infty} \left( \frac{R_*}{R_{\rm p}} \right)^{2\ell + 2} T_{\ell} (R_{\rm p})
\end{equation}
into oscillation modes, binding the star to an orbit with initial period $t_{\rm orb}$.  Here $\ell$ is a spherical harmonic modenumber and $T_{\rm \ell}<1$ is a sum over all radial modenumbers with oscillation frequencies $\omega_{n}$ (see Appendix \ref{sec:TC} for more details).  In general, the lowest order $\ell=2$ mode dominates the energy budget for mechanical oscillations.  The instability of initially linear oscillations to stochastic growth can now be understood by taking the mode phase $\phi_{n} = \omega_{\rm n} t_{\rm orb}$, and examining the per-orbit phase shift $\Delta \phi_{\rm n} = \omega_{n} \Delta t_{\rm orb}$, where $\Delta t_{\rm orb} = (3\pi/\sqrt{8})GM_\bullet \Delta E_0 / |E|^5$, and $E$ is the binding energy of the tidally captured orbit.  If $\Delta \phi > 2\pi$, as is generally the case for TC in NSCs, stochastic instability and a random walk in mode amplitude will develop \citep{IvaPap04}.
 
However, real stars possess internal dissipational mechanisms, and the presence of these qualitatively alters the evolution of a tidally captured star.  Observed circularization rates of tidally interacting stars imply an internal quality factor $Q \sim 10^{5-6}$ \citep{MeiMat05}, much lower than the $Q$ values predicted by linear dissipation theories.  This may be explicable by nonlinear mode-mode couplings; once the energy in a mode exceeds a threshold amplitude, the development of parametric instability dumps its energy into a chain of daughter modes \citep{Weinbe+12}.  These nonlinear mode-mode couplings serve as a relatively efficient source of dissipation, preventing random walks in mode amplitude from proceeding far, and leading to a steady state mode energy \citep{Lai97} comparable to the ``capture'' value after the first pericenter passage.
 
 Because $\Delta E_0$ greatly exceeds the threshold energy for parametric instability, orbital energy will be dissipated, and circularization will proceed 
on a timescale 
\begin{equation}
t_{\rm circ} = \sum_i^N t_{\rm orb}^i = t_{\rm orb} \left( \frac{E_0}{\Delta E} \right)^{3/2} \sum_{i=0}^N \frac{1}{(E_0 / \Delta E + i)^{3/2}} \label{eq:tCirc}
\end{equation}
where the $i$th orbit has a period $t_{\rm orb}^i$ and we sum over $N \equiv E_{\rm c}/\Delta E$ orbits.  Once a single star has circularized into a tight orbit of energy $E_{\rm c} = -GM_\bullet M_\star / (4R_{\rm p})$, dynamical tides will be replaced by a quasi-equilibrium tide and mass transfer will occur on the (different) dissipation timescale for equilibrium tidal friction.  We note that circularization is itself a runaway process, because the transfer of orbital energy into the star reduces the time for the next pericenter encounter.  Although the sum in Eq. \ref{eq:tCirc} can be written explicitly as a Hurwitz zeta function for $N \gg 1$, in practice $t_{\rm circ} \sim t_{\rm orb}$ provided $|E_0| \lesssim \Delta E$.

However, successful circularization requires the star to radiate at least its own binding energy.  The ratio $E_{\rm c}/E_\star = (M_\bullet/M_\star)^{2/3}/(4\lambda) \gg 1$ at late times when $M_\bullet \gg 10 M_\odot$, and $\sim 1$ when $M_\bullet \sim 10M_\odot$.   If energy is dissipated into the star faster than it can be radiated, then the star will expand in response.  This process will itself rapidly run away (as the lowest-order $\ell=2$ modes have $\Delta E_0/E_\star \propto R_\star^6$, where $E_\star = GM_\star^2/R_\star$), likely culminating in disruption \citep{IvaPap07}.  The criterion for this is whether or not the Kelvin-Helmholtz time $t_{\rm KH}=GM_\star^2/(R_\star L_\star)$ is greater than the inflation timescale
\begin{equation}
t_{\rm infl} = t_{\rm orb} \left( \frac{E_0}{\Delta E} \right)^{3/2} \sum_{i=0}^\mathcal{N} \frac{1}{(E_0 / \Delta E + i)^{3/2}}. 
\end{equation}
Here $\mathcal{N} \approx 0.1 E_\star / \Delta E \ll N$ is the number of orbits required to double $\Delta E$, assuming crudely that $R_\star^{-1} \propto E_\star + \mathcal{N}\Delta E$. 

If $t_{\rm KH} \gg t_{\rm infl}$, the star will disrupt on the inflation timescale, likely through a series of partial tidal disruptions that run away.  In practice, $t_{\rm KH}$ is generally much longer than other timescales relevant for the tidal capture problem, so that inflation is a plausible outcome.  Although $t_{\rm infl} \sim t_{\rm circ} \sim t_{\rm orb}$, $\mathcal{N} \ll N$ for all but the smallest ($\sim 1 M_\odot$) SBHs, so stars tidally captured around BHs are unlikely to circularize prior to runaway inflation \citep{Noviko+92}.  The star begins to disrupt at a characteristic semimajor axis 
\begin{equation}
a_{\rm d} \sim R_\star (M_\bullet / M_\star).
\end{equation}
In general $R_{\rm t} \ll a_{\rm d} \ll R_{\rm a}$.

The one caveat to this analysis is that the classical Kelvin-Helmholtz time may severely overestimate how long the heat requires to radiatively diffuse out of the star if it is deposited in less optically thick outer regions.  If the effective Kelvin-Helmholtz time can be shrunk by a factor $\gtrsim 10^4$, then it will shrink below $t_{\rm infl}$ and stars can be consumed by stable mass transfer \citep{DaiBla13} rather than in a sequence of runaway partial disruptions.  The final outcome is generally similar: the SBH gains an order unity fraction of the star's mass, although over a much longer timescale if tidal heating radiates away.

Crucially, the above picture assumes a tidal capture binary evolving in isolation.  Both circularization and runaway consumption can be derailed (or expedited) by orbital perturbations from other, unbound, stars.  Such a tidally captured star can be ionized due to strong perturbations (i.e. another star passing within the secondary's Hill sphere) after a time
\begin{equation}
t_{\rm ion} = \pi^{-1}\sigma^{-1}n_\star^{-1}{R}_{\rm a}^{-2} \left( \frac{M_\bullet}{M_\star} \right)^{2/3} \lesssim \frac{\sigma^3}{\pi n_{\rm c} G^2 M_\bullet^2}.
\end{equation}
Here we have taken the limit of gravitationally unfocused encounters, as the orbital apocenter $R_{\rm a} \gtrsim GM_\bullet / \sigma^2$, typically.  

On much shorter timescales than this, the angular momentum of a tidally captured star can be perturbed into weaker or stronger tidal interactions with the BH.  The timescale for an order unity change in orbital angular momentum due to two-body stellar perturbations is 
\begin{equation}
t_{\rm J} = \frac{J_{\rm orb}^2}{\langle (\Delta J)^2 \rangle} \approx \frac{\lambda}{4\pi} \frac{R_\star M_\bullet^{4/3} \sigma}{G\bar{M}_\star^2 n_{\rm c}R_{\rm a}^2 M_\star^{1/3} \ln(b_{\rm max}/b_{\rm min})}.
\end{equation}
Here $J_{\rm orb}^2 \approx 2\lambda GM_\bullet R_{\rm t}$ is the specific angular momentum of the tidally captured star, and $\langle (\Delta J)^2 \rangle$ is the specific angular momentum diffusion coefficient; we take $b_{\rm max} = r_{\rm c}$ and $b_{\rm min}=2GM_\star/\sigma^2$.
Importantly, stochastic changes in orbital angular momentum can halt mode excitation and prevent a captured star from inflating in radius, provided the pericenter moves outward.  Alternatively, a pericenter that random walks inward could see an acceleration in the inflation of its envelope.  The first of these processes is favored because of the larger available phase space away from $J=0$ in a 3D random walk, so a captured star cannot be quickly consumed if $t_{\rm J}$ is short.  

\begin{figure}
\includegraphics[width=85mm]{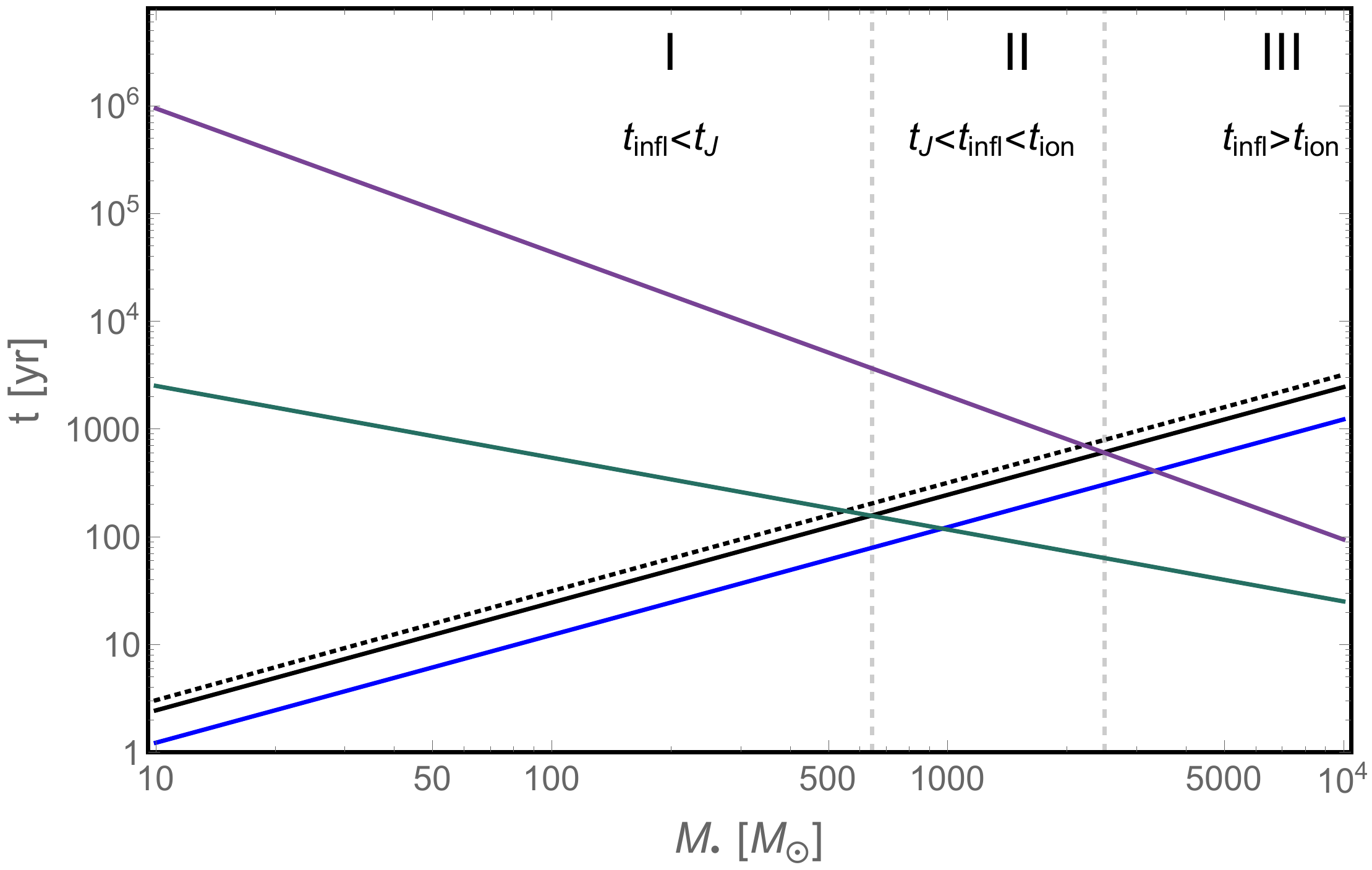}
\caption{Different timescales relevant for the evolution of tidally captured stars.  The shortest is often the orbital time, $t_{\rm orb}$, shown in blue, although for large values of $M_\bullet$ it is the angular momentum relaxation time $t_{\rm J}$, shown in green.  The time for the star to enter a runaway inflation regime, $t_{\rm infl}$ (black solid), and the time for orbital energy loss to circularize the orbit of the captured star, $t_{\rm circ}$ (black dotted) are each small multiples of $t_{\rm orb}$.  Because $t_{\rm infl} < t_{\rm circ}$, only the former process is relevant.  The binary ionization timescale $t_{\rm ion}$ is shown in purple, and is generally greater than $t_{\rm J}$.  The longest timescale in the problem (not shown) is the Kelvin-Helmholtz time for a captured star to radiate dissipated mode energy.  When $t_{\rm infl} < t_{\rm J}$ (regime I), captured stars are efficiently consumed via a series of runaway partial tidal disruptions.  When $t_{\rm J} < t_{\rm infl} < t_{\rm ion}$ (regime II), stars will quickly random walk to larger pericenters where tidal forces are initially irrelevant, but may still inflate and be disrupted through collisions with other tidally captured stars.  When $t_{\rm ion} < t_{\rm infl}$ (regime III), captured stars cannot inflate significantly before being ejected or swapped out by encounters with unbound stars.  In this diagram, $\sigma = 60~{\rm km~s}^{-1}$, $n_{\rm c} = 10^8 ~{\rm pc}^{-3}$, and we have used a Kroupa IMF truncated at a maximum mass of $1M_\odot$.}
\label{fig:TCTimes}
\end{figure}

We illustrate the relative timescales for these processes in Fig. \ref{fig:TCTimes}.  When $M_\bullet$ is small, the shortest timescale is the orbital period $t_{\rm orb}$, followed by $t_{\rm infl}$.  When $t_{\rm infl} < t_{\rm J}$, thermalization of mode energy will result in repeated partial disruptions and eventually the full disruption of the star; when $t_{\rm J} < t_{\rm infl}$, this can still occur, but most stars will also random walk to large, non-interacting values of angular momentum, a possibility we address momentarily.  In general, the Kelvin-Helmholtz time is too long to affect the evolution of a tidally captured star, and while $t_{\rm circ} \sim t_{\rm infl}$, the fact that $\mathcal{N} \ll N$ prevents circularization and stable mass transfer from occurring.  Generally, $t_{\rm ion} \gg t_{\rm J}$.

The preceding discussion has focused on the tidal capture of a single star by a SBH; however, once the black hole has entered the runaway growth regime, it will begin tidally capturing stars at intervals less than $t_{\rm infl}$ (especially if angular momentum relaxation delays stellar consumption).  At this point multi-body interactions will become important.  If the BH is orbited by $N_{\rm TC}$ tidally captured stars that have been unable to circularize or expand through mode thermalization (i.e. have random walked to slightly larger pericenter values), most will possess orbital apocenters $R_{\rm a} \sim GM_\bullet / \sigma^2$, and they will map out a 3D density profile 
\begin{equation}
n_{\rm TC}(R) = \frac{3N_{\rm TC}}{8\pi R_{\rm a}^3} \left(\frac{R}{R_{\rm a}} \right)^{-3/2}.
\end{equation}
In general, this ``mini-cusp'' will have a much higher density than the surrounding star cluster, and will be prone to direct physical collisions between the stars it contains.  These collisions will destroy the involved stars if they occur at a radius $R \lesssim a_{\rm d}$, which is also the radius at which gravitational focusing of encounters between mini-cusp stars becomes important.  Destructive collisions primarily happen at orbital pericenter at a per-star rate
\begin{align}
\dot{N}_{\rm des} \sim& \frac{3N_\star R_\star^2 \sigma^7}{8G^3M_\bullet^3}\\
\sim& 4\times 10^{-6}~{\rm yr}^{-1} \frac{N_\star}{10} \left(\frac{R_\star}{R_\odot} \right)^2 \left(\frac{\sigma}{40~{\rm km}/{\rm s}} \right)^7 \left(\frac{M_\bullet}{10M_\odot} \right)^{-3}.  \notag
\end{align}
In contrast, weaker (soft) coagulative collisions happen preferentially at apocenter, with a per-star rate
\begin{align}
\dot{N}_{\rm coag} \sim& \frac{3N_\star M_\star R_\star \sigma^5}{4G^2M_\bullet^3}\\
\sim& 10^{-3}~{\rm yr}^{-1} \frac{N_\star}{10} \frac{M_\star}{M_\odot} \frac{R_\star}{R_\odot} \left(\frac{\sigma}{40~{\rm km}/{\rm s}} \right)^5 \left(\frac{M_\bullet}{10M_\odot} \right)^{-3}.  \notag
\end{align}
Generally, coagulative collisions happen at rates orders of magnitude higher than destructive ones, which we hereafter neglect.  These stellar mergers will nonetheless deposit huge amounts of kinetic energy into the merger product; much of this will thermalize promptly due to shocks, inflating the star to the point where it can likely be tidally disrupted on a subsequent pericenter passage, or grow to a size where it could engulf the central BH.

The few-body dynamics of these tidal capture mini-cusps are likely complex, as scalar resonant relaxation of stellar orbits should be suppressed by GR apsidal precession \citep{HopAle06}, even vector resonant relaxation may be likewise suppressed if the BH is spinning \citep{MerVas12}, and the number of stars may be too few for standard two-body relaxation approximations to hold.  Nonetheless, basic density considerations indicate that direct collisions should be frequent, and the aftermath of these will be inflated stars that are easily disrupted by the BH.  Once the BH has grown to a very large size, typically $\gtrsim 10^3 M_\odot$, ionization of tidally captured stars in encounters with unbound stars will become frequent ($t_{\rm ion} < t_{\rm orb} < t_{\rm infl}$).  At this point tidal capture may turn off as a growth channel, leaving only full tidal disruptions to grow the BH.

We therefore conclude that tidal capture events generally result in delayed tidal disruptions due to inflation of the captured star: this occurs either as a result of runaway mode excitation \citep[in analogy to][]{LiLoe13}, or following the thermalization of direct soft collision kinetic energy.  These delayed tidal interactions might be quite different from standard TDEs in that they involve interaction of the BH with stars that have radially inflated and are possibly more massive than normal stars.  We will assume that the end point of such events is the consumption of the bulk of the mass of the normal star by the SBH.  This will likely occur at super-Eddington rates while the black hole is small.  Such super-Eddington accretion disks may lose some mass in outflows, though the exact amount is still debated in the simulation literature \citep[for example, find $\approx 30\%$ of the inflowing mass lost in a wind]{Jiang+14}.  We neglect mass loss in super-Eddington outflows, but emphasize that it could become important for SBH growth if it reaches an order unity fraction of the mass inflow rate.

\section{Phase-space diffusion and few-body dynamics}
\label{sec:numerics}

\begin{figure}
\includegraphics[width=85mm]{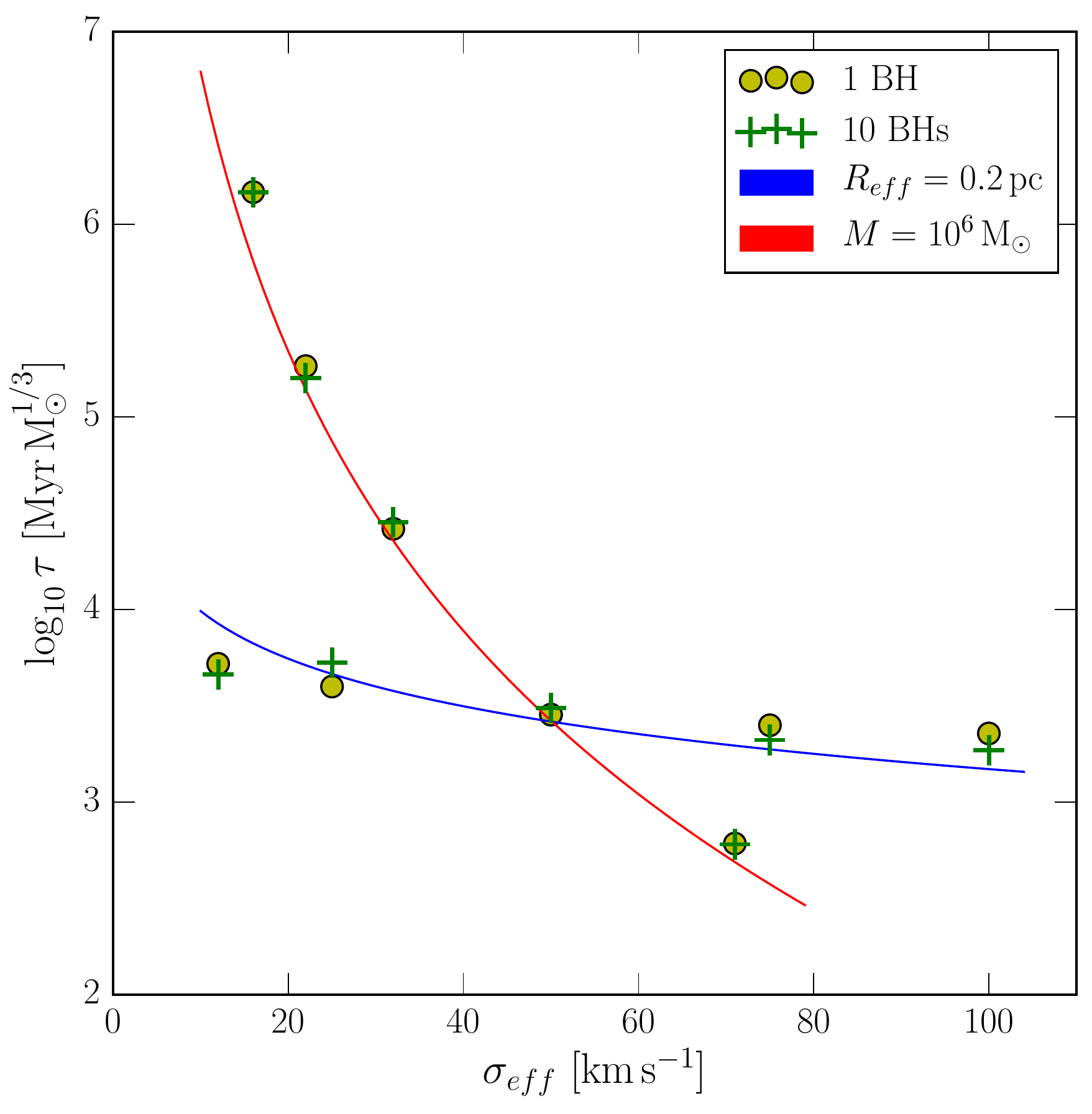}
\caption{A comparison between analytic predictions (lines), one-body integrations (circles), and direct few-body integrations (crosses).  For all three types of calculations, we show the runaway parameter $\tau$ for SBHs in unconcentrated Plummer spheres, where $\dot{M}_\bullet \equiv M_\bullet^{4/3}/\tau$.  For the analytic predictions, we use Eq.~\ref{eq:TCRate}, while we measure this from simulation data by following the time evolution of the most massive surviving black hole.  The general agreement between analytic theory and few-body integrations (which all employ 10 SBHs) validates our single-body analytic growth rate approximation for dense star systems.}
\label{fig:fewBody}
\end{figure}

The analytic description of runaway SBH growth presented in the prior section neglects some basic dynamical processes that are difficult to account for in a closed form. For this reason, we have performed two sets of simple numerical simulations to test the validity of our analytical prescriptions in a more realistic, dynamical setting.

The first set are one-body orbital integrations of single SBHs in an analytic background potential representing an NSC. The SBH's diffusion through phase space is treated analytically through calculation of the diffusion coefficients (e.g., \citealt{BinTre08}) at each time step and respective modification of its velocity vector. The background potential is taken as static and only evolves if the SBH ``feeds'' on it, i.e., by (probabilistically) capturing or disrupting a star, which reduces the mass of the NSC. Moreover, the scale radius of the background potential can expand due to dynamical heating from the SBH. These highly idealized simulations are valuable for testing our approximate ``$n\Sigma v$'' estimates, and the influence of phase-space diffusion on the growth rates of the SBHs.

The second set of simulations are few-body integrations of a small number of SBHs in the same type of background potential. These simulations are significantly more realistic in that they include mergers of SBHs and their dynamical ejections through gravitational-wave recoils or few-body interactions. They provide a more physical test of the analytic model of \S \ref{sec:clusterObs}. Comparisons of our analytic prescriptions with results from both sets of simulations are presented in the following. For a detailed description of the numerical simulations see Appendix~\ref{sec:simulations}.

\subsection{Simulation results}

In Fig.~\ref{fig:fewBody}, we show how analytic, single-body and few-body estimates of the growth rates of SBHs in NSCs compare. For this purpose, we define a ``runaway time scale'' $\tau$ such that Eq.~\ref{eq:TCRate} can be written as
\begin{equation}
\dot{M}_\bullet = M_\bullet^{4/3}/\tau,
\end{equation}
with $\tau$ having dimension Myr\,M$_\odot^{1/3}$. Solving this equation for $M_\bullet$ gives us the mass of the SBH with time
\begin{equation}
M_\bullet(t) = \left(-\frac{t-t_0}{3\tau} + M_{\bullet,0}^{-1/3}\right)^{-3}.
\end{equation}
The runaway time scale can therefore be calculated from any two snapshots of a numerical simulation using
\begin{equation}
\tau = -3\frac{M_\bullet(t)^{-1/3}-M_{\bullet,0}^{-1/3}}{t-t_0}.
\end{equation}
As we can see in Fig.~\ref{fig:fewBody}, the deviations between the analytic predictions and the numerical simulations is small across a wide range of NSC velocity dispersions, $\sigma_{eff}$, relevant for our study. Moreover, few-body dynamical effects, such as ejections and mergers, have a low but highly stochastic effect on the runaway time scales.  In general, it appears that our analytic prescriptions reliably describe the rate at which SBHs gain mass from interactions with unbound stars in both the one-body and few-body regimes.  As noted earlier, these prescriptions likely break down if a self-gravitating, Spitzer-unstable subcluster of SBHs can form, but such a phase of cluster evolution is likely short-lived, eventually culminating in the few-body regime probed by our simulations.

In reality, the assumptions underlying the initial phase of super-exponential growth will break down once the an SBH seed has grown large enough to dominate the potential surrounding it, as we quantify in the following section.  The importance of multi-SBH effects (which already appears to be small, based on the results in Fig. \ref{fig:fewBody}) will also diminish, as remaining SBHs are efficiently ejected from the centers of clusters with a massive IMBH \citep{Leigh+14b}.

\section{Outcomes of Runaway Growth}
\label{sec:outcomes2}

Although two effects (binary ionization, angular momentum diffusion) will begin impeding tidal capture runaways once the SBH has grown to a mass $M_\bullet \sim 10^{2-3} M_\odot$, neither of these can cleanly terminate the runaway, as each becomes much smaller if the initial pericenter of capture is moved somewhat inward.  They can therefore be thought of as modest reductions in the tidal capture cross-section.  Even if tidal capture were completely deactivated, IMBH growth through tidal disruptions would continue so long as debris streams were not deflected by perturbations from cluster stars.  Thus tidal disruption continues to grow the black hole provided
\begin{align}
n_\star \lesssim& 2^{-15/4}\beta^{-3/4}R_\star^{-3} \left(\frac{M_\bullet}{M_\star} \right)^{-3/2} \\
\lesssim & 6 \times 10^9 ~{\rm pc}^{-3} \beta^{3/4} m_\star^{3/2} r_\star^{-3} \left(\frac{M_\bullet}{10^8 M_\odot} \right)^{-3/2} \notag.
\end{align} 
Tidal disruptions therefore can continue to grow the IMBH up to very large masses in any realistic star cluster, and the factor of $\lambda$ reduction in their cross-section (relative to tidal capture) will matter little in the runaway growth regime.

Since accretion and binary physics seem unlikely to deactivate tidal growth channels, the termination of the runaway must instead have a stellar dynamical origin.  In this section, we identify three stages of black hole growth through star capture that correspond to three different BH mass ranges.  The first stage is the one we have considered so far in this paper, where the SBH drifts through a cluster core whose stars have not yet been depleted through tidal capture/disruption.  In this stage, the rate of SBH growth is $\dot{M}_\bullet = n_\star M_\star \Sigma \sigma \propto M_\bullet^{4/3}$.  During this (and later) stages, the black hole moves in the almost flat potential of the cluster core.  Its orbit can be thought of as Brownian motion: damped (by dynamical friction) and stochastically driven (by diffusive scatterings) movements in a simple harmonic oscillator \citep{BahWol76, Chatte+02a, Chatte+02b}.  The RMS value of the black hole's orbital radius is 
\begin{equation}
\langle r_{\rm w}^2 \rangle \approx \frac{M_\star}{M_\bullet} r_{\rm c}^2.
\end{equation}
There is an order unity prefactor in $r_{\rm w}$ which depends on the details of the star cluster distribution function; we neglect it for simplicity.  For the remainder of this section we take $M_\star$ to be the average stellar mass in the cluster's present day stellar mass function.  Because even the gravitationally focused tidal capture radius $\sqrt{2R_{\rm TC}GM_\bullet/\sigma^2} \ll r_{\rm c}$, it is unlikely that the BH can ever deplete its ``wandersphere.''  However, as the BH grows, very quickly the wandersphere will recede inside the influence radius $r_\bullet \equiv GM_\bullet/\sigma_{\rm c}^2$.  

Once $r_{\bullet} > r_{\rm w}$, the BH is no longer swimming through a sea of stars but is instead orbited by a bound minicluster and a much larger cluster of unbound stars.  At this point the rate of tidal consumption may no longer be well approximated by an $n\Sigma v$ calculation.  The end of the first stage of SBH growth occurs once it reaches a mass
\begin{equation}
M_{\rm w} = M_\star \left( \frac{6(\pi^2/8-1)}{\pi} \right)^{2/3} \left(\frac{M_{\rm tot}r_{\rm c}}{M_\star r_{\rm h}} \right)^{2/3} M_\star
\end{equation}
which is generally small, $M_{\rm w}\approx M_{\rm c}(M_\star / M_{\rm tot})^{1/3}\sim 10^{1-3} M_\odot$.  If large stellar mass black holes \citep[e.g. if star formation occurs at low metallicity;][]{Belczy+10} exist at the birth of a small or highly concentrated cluster, the first growth stage may be bypassed all together.

In the second stage of black hole growth, the rate of star consumption is still confined to the NSC core but is now diffusion-limited, as in the classical loss cone problem \citep{FraRee76, LigSha77, CohKul78}.  We estimate diffusion-limited consumption rates using the simplified formalism of \citet{SyeUlm99}, who find that the stellar consumption rate is $\dot{N} = \dot{N}_< + \dot{N}_>$,
where 
\begin{equation}
\dot{N}_< = 4\pi \int^{r_{\rm crit}}_0 \frac{G^2 \rho^2 r^2}{k \sigma^3}\frac{\ln\Lambda}{\ln(2/\theta_{\rm LC})}\frac{\langle M_\star^2 \rangle}{M_\star^2}{\rm d}r
\end{equation}
and
\begin{equation}
\dot{N}_> = 4\pi \int_{r_{\rm crit}}^\infty \frac{GM_\bullet R_{\rm t} \rho}{r M_\star \sigma} {\rm d}r
\end{equation}
are the stellar consumption rates in the empty and full loss cone regimes, respectively.  We note that $\sigma(r) = \sqrt{\sigma_{\rm c}^2+GM_\bullet/r}$ in these integrals; the angular size of the loss cone $\theta_{\rm LC}^2 \equiv GM_\bullet R_{\rm t}/(\sigma^2r^2)$.  We have used the second moment of the present day stellar mass function, $\langle M_\star^2 \rangle$.  The critical radius $r_{\rm crit}$ is the location where the per-star consumption rate is equally partitioned between the full and empty loss cone regimes.  It can be found by solving the equation
\begin{equation}
r_{\rm crit} = 2\pi^2 k \frac{\ln(2/\theta_{\rm LC})}{\ln\Lambda} \frac{r_{\rm h}\sigma^2(r_{\rm crit})}{GM_{\rm tot}} \frac{M_\bullet M_\star}{\langle M_\star^2 \rangle} R_{\rm t},
\end{equation}
for $r_{\rm crit}$ (which cannot be done in closed form, as $\theta_{\rm LC}$ is a function of $r_{\rm crit}$).  The dimensionless constant $k \approx 0.34$ is a numerical prefactor from the relaxation time (Eq. \ref{eq:trelax}).

During the second stage of BH growth, $r_{\rm w} < r_\bullet < r_{\rm c}$.  Initially, $\dot{N}_> \gg \dot{N}_<$ and most of the consumed stars come from the full loss cone regime; the total consumption rate here is comparable to our earlier $n\Sigma v$ estimate, and still scales as $\dot{M}_\bullet \propto M_\bullet^{4/3}$.  As the black hole grows further, the two regimes of loss cone fueling become comparable in magnitude, slowing the runaway.  However, the IMBH growth rate remains roughly exponential until the entire cluster core has been consumed, at which point $\dot{N}_>$ becomes subdominant and $\dot{N}$ quickly decelerates.

Neglecting the full loss cone regime, we can find an asymptotic late time (large $M_\bullet$) solution.  In the empty loss cone regime, most of the stellar flux into the loss cone originates from the smaller of $r_{\rm crit}$ and $r_\bullet$.  Generally, $r_{\rm crit} < r_\bullet$, in which case we find that $\dot{N} \approx \dot{N}_< \propto M_\bullet^{-11/12}$ (here we have also neglected the weak dependence of $\dot{N}$ on logarithmic functions of $M_\bullet$).  More precisely, we find under this approximation (and assuming that $r_{\rm c} \ll r_{\rm h}$) that
\begin{align}
\dot{N}_< \approx& \frac{2^{5/4}}{\pi^{5/2}k^{3/4}}\left( \frac{\ln \Lambda}{\ln(2/\theta_{\rm LC})} \right)^{3/4} \left(\frac{M_\star}{\langle M_\star^2 \rangle} \right)^{1/4} \left( \frac{GM_{\rm tot}}{r_{\rm h}} \right)^{7/4} \\
& \times G^{-5/4}R_\star^{1/4}M_\star^{-1/12} M_\bullet^{-11/12} .\notag
\end{align}
Star capture predominantly in the empty loss cone regime (after the IMBH has reached a mass $M_\bullet \sim M_{\rm c}$) marks the third and final phase of black hole growth in NSCs.  Integrating $\dot{M}_\bullet = \dot{N}_<(M_\bullet)M_\star$, we see that the black hole mass converges to a late-time value that only depends strongly on $\sigma$:
\begin{align}
M_{\rm sat} \approx & 1.6 \left(\frac{\ln\Lambda}{\ln(2/\theta_{\rm LC})} \right)^{9/23} \left(\frac{\sigma_{\rm c}^{7/2}t}{V_\star^{5/2}R_\star} \left(\frac{M_\star^2}{\langle M_\star^2 \rangle} \right)^{1/4} \right)^{12/23} \label{eq:MSat}  \\
\approx & 6 \times 10^5 M_\odot ~ \sigma_{40}^{42/23} \left( \frac{t}{10^{10}~{\rm yr}} \right)^{12/23} . \notag
\end{align}
In the second of these equations we have taken typical stellar masses and radii from a Kroupa IMF, and have used $M_\bullet = 10^5 M_\odot$ inside the logarithmic terms (though the final prefactor depends only weakly on this choice).  We have also used the shorthand $V_\star \equiv \sqrt{GM_\star / R_\star}$.

This ``saturation mass'' is reached at late times after the entire cluster core has been consumed, provided the IMBH does not grow through other means (e.g. AGN activity).  Our model therefore makes clear predictions for the mass of IMBHs grown through the mechanisms in this paper, predictions which to first order depend only on $\sigma$.  We note that because of the approximate one-to-one correspondence between $\sigma$ and the velocity dispersion of the host galaxy \citep{Leigh+15}, this is in principle testable even when the NSC cannot be resolved.  We compare these predictions to observations in \S \ref{sec:observations}.

When $M_\bullet$ is relatively small, $r_{\rm crit} < r_\bullet$ and the above approximations are valid for the saturation mass; in this regime $r_{\rm crit}/r_\bullet \sim (\sigma/V_\star)(M_\bullet / M_\star)^{1/6}$.  For high-$\sigma$ NSCs, it is possible for SMBHs in the diffusion-limited growth regime to grow to the point where $r_\bullet < r_{\rm crit}$, in which case the asymptotic growth rate formula changes somewhat.  In this limit we find a similar but simpler version of Eq. \ref{eq:MSat}, 
\begin{align}
M_{\rm sat}' \approx& 2.9 \left(\frac{\langle M_\star^2 \rangle}{M_\star^2} \right)^{1/2} \left(\frac{\ln\Lambda}{\ln(2/\theta_{\rm LC})} \right)^{1/2}  \left(\frac{M_\star \sigma_{\rm c} t}{G} \right)^{1/2} \\
\approx& 1.7\times 10^6 M_\odot \sigma_{40}^{3/2} \left(\frac{t}{10^{10}~{\rm yr}} \right)^{1/2}.   \notag
\end{align}
This extremely simple formula ($M_{\rm sat}' \sim \sqrt{M_\star \sigma_{\rm c}^3t/G}$) represents the final saturation mass for an IMBH growing through tidal capture of stars in a cluster with $M_{\rm c} < M_\bullet < M_{\rm tot}$, but in practice it is only reached for high-$\sigma$ systems; lower ones saturate at the similar mass given by Eq. \ref{eq:MSat}, which has slight differences in the prefactor and scaling exponents.

\begin{figure}
\includegraphics[width=85mm]{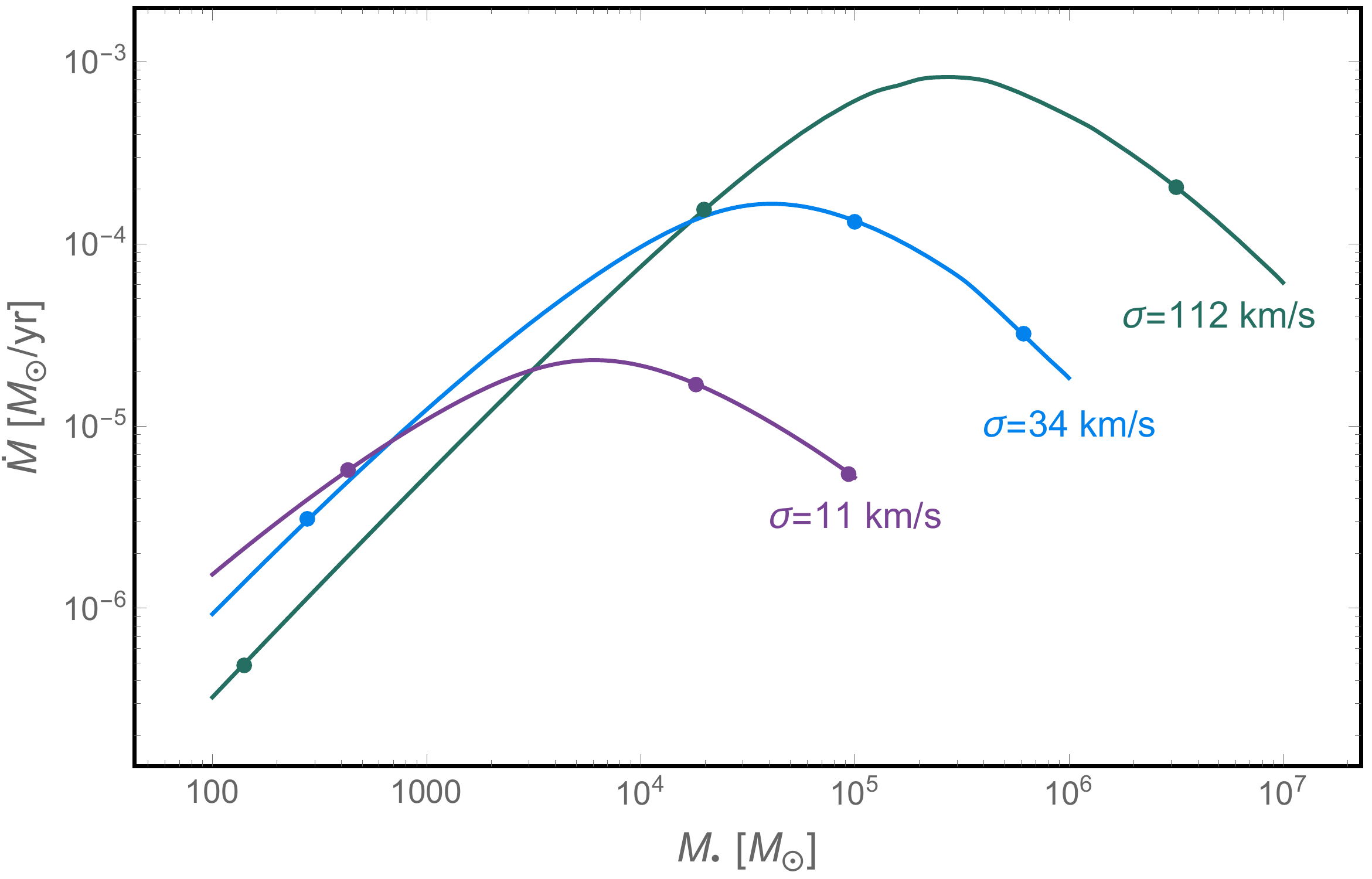}
\caption{The mass growth histories of SBH seeds and developing IMBHs in our model; we plot $\dot{M}_\bullet$ in as a function of black hole mass for clusters with initial central densities $\rho_{\rm c}=10^8~{\rm pc}^{-3}$ (i.e. on the cusp of runaway growth for a $M_\bullet =10 M_\odot$ seed.  The purple, blue, and green curves have initial core radii $r_{\rm c}=0.01~{\rm pc}$, $r_{\rm c}=0.03~{\rm pc}$, and $r_{\rm c}=0.1~{\rm pc}$, respectively, corresponding to different velocity dispersions which are labeled in the plot.  The three dots along each curve label $M_\bullet$ at times $10^8~{\rm yr}$, $10^9~{\rm yr}$, and $10^{10}~{\rm yr}$ after the SBH enters the cluster core with a mass of $M_\bullet = 100M_\odot$.  }
\label{fig:MDotEdd}
\end{figure}

The overall growth history of example IMBHs is shown in Fig. \ref{fig:MDotEdd}.  Notably, SBHs that are able to enter the runaway regime within a Hubble time often require periods of super-Eddington growth; throughout this paper we assume that these rates of mass inflow are possible, but we emphasize here that such an assumption is still an open question in the accretion literature \citep{Jiang+14, Sadows+15}.

Another notable feature of our scenario is that SMBH seeds formed through tidal capture runaways are likely born with negligible spin.  Although the onset of the runaway may involve tidal captures of the innermost stars, once diffusion-limited growth begins, it will come primarily from stars on radial orbits taken from more distant regions of the star cluster.  Even if the cluster is endowed with net rotation, such rotation is unlikely to be imprinted on the almost radial orbits vulnerable to tidal capture or disruption.  In contrast, SMBHs that have grown substantially through comparable-mass mergers or through accretion from gas inflows on larger scales will generally approach larger values of spin \citep{BerVol08}, though this can be prevented if mass accretion is dominated by short-lived episodes of gas inflow from random directions \citep{KinPri06}.

\section{Comparison with Observations}
\label{sec:observations}

So far, we have seen that clusters of sufficiently high central density and velocity dispersion will enter a runaway (super-exponential) phase of BH growth.  This runaway regime begins slowing as the loss cone depletes, and then putters out into a slower diffusion-limited growth rate once the BH has eaten the entire core of its host cluster.  Even at this more modest diffusion-limited rate, however, BHs can grow significantly over a Hubble time by tidal capture and disruption.

In this section, we compare the predictions of our model to the observed demographics of NSCs, and to signatures of massive black holes in galactic nuclei.  These two data sets present several tests of our model, best formulated as questions.  Do we see NSCs unstable to a tidal capture runaway that lack massive central black holes?  And do we see a SMBH mass distribution in galactic nuclei that falsifies our model of SMBH seed formation?

\subsection{Observed NSCs}

\begin{figure}
\includegraphics[width=85mm]{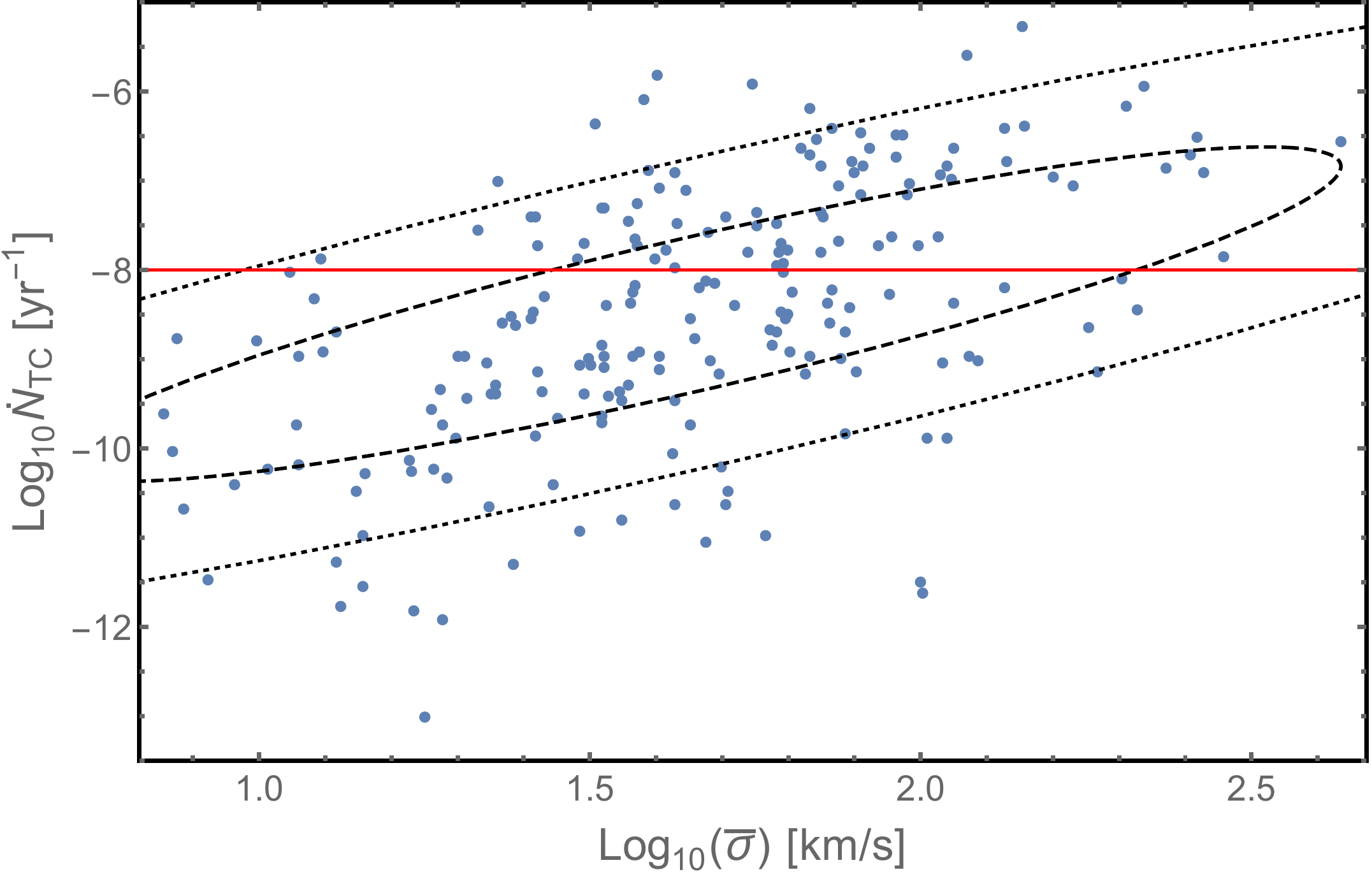}
\caption{Tidal capture rates, $\dot{N}_{\rm TC}$, plotted against 1D NSC velocity dispersion $\bar{\sigma}$.  In all of these calculations $\dot{N}_{\rm TC}$ is presented for $M_\bullet = 10M_\odot$.  The dashed and dotted black ellipses show the $1-$ and $2-\sigma$ contours.  Notably, many of the more massive NSCs are above the solid red line and have tidal capture runaway timescales $< t_{\rm H}$, indicating that they may already harbor unresolved massive black holes.  Almost no NSCs exist in the runaway regime that have $\bar{\sigma}<30~{\rm km~s}^{-1}$.}
\label{fig:TCRates}
\end{figure}

Using the simple analytical estimates for tidal capture rates that we derived in \S \ref{sec:clusterObs} and validated in \S \ref{sec:numerics}, we can now examine observed NSCs to determine their tidal capture runaway timescales.  Fig. \ref{fig:TCRates} plots the tidal capture rate $\dot{N}_{\rm TC}$ as a function of 1D cluster velocity dispersion $\bar{\sigma}$ (as before, this is estimated using the fitted cluster mass and radius) for an SBH with $M_\bullet = 10 M_\odot$, assuming that there is no IMBH or SMBH to modify the central potential.  We note that estimates of the tidal capture rate are uncertain by one order of magnitude because of a factor $\approx 3$ uncertainty in the fitted cluster concentration parameter.  Specifically, all clusters in this sample were fit to a grid of King models, but this grid was coarsely sampled in the dimension of $r_{\rm c}$: the only possible concentrations were $5, 15, 30,$ and $100$ (Georgiev, private communication).

Many of the NSCs in Fig. \ref{fig:TCRates} have tidal capture rates $\dot{N}_{\rm TC}\gtrsim 10^{-8}~{\rm yr}^{-1}$, indicating that they are formally unstable to runaway SBH growth through tidal capture.  These NSCs generally have $\bar{\sigma} \gtrsim 30~{\rm km~s}^{-1}$, and in some cases have truly enormous TC rates (due to their high central densities).  We highlight here two systematic uncertainties that may cause substantial error in our estimates of central density and derived quantities such as $\dot{N}_{\rm TC}$: first, the grid of $r_{\rm c}$ only extends up to concentrations $C=100$, meaning that many clusters may be even more concentrated and have even higher $\dot{N}_{\rm TC}$.  Secondly, the estimates of cluster mass $M_{\rm tot}$ assume a constant mass-to-light ratio, and do not account for color gradients due to dynamical or primordial mass segregation.  Both of these effects concentrate surface brightness in the center of the NSC and cause an {\it overestimation} of cluster concentration $C$.  Observations of individual NSCs often find color gradients, but the direction of the gradient can vary from cluster to cluster \citep{KorMcC93, Matthe+99, Carson+15}.  A careful resolution of these two uncertainties is important, but will have to wait for future observational work.

We also can compare our predictions to the smaller number of galaxies which have both NSC mass measurements (or upper limits) and dynamical SMBH mass measurements (or upper limits).  In Fig. \ref{fig:NSCDichotomy2} we show our prediction for saturation mass $M_{\rm sat}$ for massive black holes that have been growing through star capture for a Hubble time.  We plot $M_{\rm sat}$, and measured NSC/SMBH masses, against the host galaxy's effective dispersion $\sigma_{\rm eff}$, which is comparable to the NSC internal velocity dispersion \citep{Leigh+15}.  We find that the smallest SMBHs with dynamically measured masses ($10^6M_\odot \lesssim M_\bullet \lesssim 10^7 M_\odot$) are within a factor of a few of $M_{\rm sat}$.  In galaxies with a larger $\sigma_{\rm eff}$, SMBH masses are much larger than $M_{\rm sat}$, implying that other mechanisms dominate the growth of these SMBHs, in agreement with the Soltan argument.  In smaller galaxies with $\sigma_{\rm eff} \lesssim 40~{\rm km~s}^{-1}$, dynamical mass measurements only put upper limits on the presence of an SMBH, and these upper limits are generally well below the predicted saturation mass.  This implies that, as expected, a tidal capture runaway has not occurred in these low velocity dispersion systems.

This low-$\sigma_{\rm eff}$ result is in reasonable agreement with our prediction that NSCs must have a minimum central density in order to begin growing SBHs through runaway star capture.  As we see in Fig. \ref{fig:TCRates}, it is only NSCs with $\bar{\sigma} \gtrsim 30~{\rm km~s}^{-1}$ that possess high enough central density to enter the runaway regime within a Hubble time.  This impressive pair of observational coincidences - the lack of SMBHs in galaxies with $\bar{\sigma} \lesssim 40~{\rm km~s}^{-1}$ {\it and} the unfavorably low central densities of NSCs in galaxies with $\bar{\sigma} \lesssim 30~{\rm km~s}^{-1}$ - is a nontrivial piece of circumstantial evidence in favor of our model for the birth of massive black holes in galaxies of moderate velocity dispersion.

\subsection{Observed IMBHs}
The evidence for IMBHs in the nuclei of dwarf or other galaxies is mixed, but these observations are a crucial discriminant between models of SMBH seed formation, which predict very different minimum IMBH masses: $\sim 10^{2-3}M_\odot$ for Pop III supernovae, $\sim 10^{3-4} M_\odot$ for traditional ``runaway collision'' scenarios, and $\sim 10^{5-6}M_\odot$ for direct collapse of high-$z$ minihalos.  In our scenario, we expect a distribution of black hole masses; if the runaway time is $\ll t_{\rm H}$, then typically $M_\bullet \sim M_{\rm sat}(t_{\rm H}, \sigma)$.  It is possible for our scenario to produce smaller IMBHs in clusters where the runaway time is $\sim t_{\rm H}$, but because this requires some tuning we expect these to be rare.  Our model therefore predicts a general minimum IMBH mass at $M_{\rm sat}(t_{\rm H}, \sigma)$, and the distribution or even existence of small IMBHs offers a promising way to falsify our predictions.

Due to the difficulty of dynamical mass measurements in dwarf galaxies, the most common avenues for IMBH detection are indirect.  Searches for AGN in dwarf galaxies have returned many hundreds of promising candidates \citep{GreHo04, Reines+13}.  The smallest of these have lower mass limits $M_\bullet \gtrsim 10^{3-4} M_\odot$ based on luminosity arguments \citep{Moran+14}.  A smaller number of dwarf galaxies have dynamical mass estimates for central BHs; one particularly exciting recent result found a $M_\bullet \approx 5\times10^4 M_\odot$ \citep{Baldas+15}.  We note that other, future, avenues for IMBH detection do exist.  For example, the halo of the Milky Way galaxy may host many IMBHs left over from the interplay between hierarchical structure formation scenarios and GW recoil during IMBH-IMBH mergers \citep{OLeLoe09, Merrit+09}.  While these wandering IMBHs are in principle detectable by the hypercompact stellar systems that surround them, none have been found to date \citep{OLeLoe12}.


The existence of a large population of IMBHs with $M_\bullet \ll M_{\rm sat}$ would not be expected in the context of our model, and there is not yet clear evidence that such a population exists.  

\subsection{Tidal Disruption Rates}

A final consistency check for our model is the rate of stellar tidal disruption events.  From Fig. \ref{fig:MDotEdd} we see that an NSC hosting a massive black hole of mass $M_\bullet = 10^5 M_\odot$ should be tidally disrupting stars at a rate $10^{-5}~{\rm yr}^{-1} \lesssim \dot{N} \lesssim 10^{-3}~{\rm yr}^{-1}$, depending on the host NSC $\sigma$.  These numbers are roughly consistent with empirically calculated rates of stellar tidal disruption \citep{StoMet14}, although we caution that these empirical TDE rates are based on extrapolating directly determined TDE rates in larger galaxies down to smaller masses (where surface brightness profiles cannot be resolved on the relevant scales).  

A more useful constraint is the volumetric tidal disruption rate, as this is dominated by the smallest galaxies which host central massive BHs \citep{WanMer04, StoMet14}.  Although our current sample of observed tidal disruption flares is small and suffers from selection effects (roughly a dozen optically selected events, and a similar number detected through soft X-ray emission), it appears that the volumetric event rate inferred from observations is at the extreme low end of theoretical predictions \citep{vanFar14, Decker+16}, which may imply an absence of low mass ($M_\bullet \lesssim 10^6 M_\odot$) black holes in the low redshift universe \citep{StoMet14}.  Although individual TDEs may arise from smaller IMBHs\footnote{For example, an X-ray selected TDE flare was recently found in a small dwarf galaxy.  If we infer the BH mass from standard galaxy scaling relations, we find $10^{5.1} \lesssim M_\bullet/M_\odot \lesssim 10^{5.7}$ \citep{Maksym+14}.  }, a high occupation fraction of IMBHs in dwarf galaxies would be hard to reconcile with the observed low volumetric TDE rate.

We note here that past works have shown that star capture can contribute significantly both to the mass growth of small SMBHs \citep{MagTre99}, and to the lower end of the X-ray AGN luminosity function \citep{Milosa+06}.  However, these past arguments have been based on present-day observations of low-mass SMBHs and their host galaxies.  The most novel contribution of this paper is the self-consistent evolutionary picture we have presented, which shows how these low-mass SMBHs must emerge from NSCs above a certain $\rho_{\rm c}$ and $\sigma$ threshold.

\section{Conclusions}
We have shown that stellar mass black holes in nuclear star clusters will undergo runaway growth into IMBHs through tidal interactions with cluster stars.  At early times ($M_\bullet \lesssim 100 M_\odot$) this process is super-exponential and is dominated by unbound stars.  At intermediate times ($100 M_\odot \lesssim M_\bullet \lesssim M_{\rm c}$) black hole growth is dominated by bound stars from the full loss cone regime and is initially super-exponential, but slows as the cluster core depletes.  At later times, the IMBH has eaten the entire mass of the cluster core and its diffusion-limited growth slows substantially, saturating at a finite value $M_{\rm sat}\propto \sigma^{42/23}t^{12/23} \sim 10^{5-6}M_\odot$, as derived in Eq. \ref{eq:MSat} (or, for high-$\sigma$ systems, at a value $M_{\rm sat}' \propto \sigma^{3/2} t^{1/2}$).  Thus the final mass of a massive black hole grown solely through tidal interactions with stars in an NSC depends primarily on the NSC velocity dispersion $\sigma$.


We emphasize that theoretical uncertainties remain concerning both the processes of tidal disruption and tidal capture.  In the simplest models for tidal disruption events, exactly $1/2$ of the stellar mass is bound to and accreted by the black hole \citep{Rees88}; however, it is possible that the dissipative processes required to form an accretion disk may unbind a large fraction of the initially bound gas \citep{Ayal+00, MetSto15}.  Whether this occurs in practice is an open question in TDE research.  Even if this type of ``leakiness'' deactivates tidal disruption as a growth channel, tidal captures will continue to grow the black hole.  While this channel may deactivate at higher ($\sim 10^{3-4} M_\odot$) IMBH masses, this is not certain, and the buildup of a tight cluster of bound stars may lead to a cascade of soft collisions, inflating the stars and allowing them to be consumed.  We also note that both channels imply that a SBH seed that grows into an IMBH or SMBH in a Hubble time will likely pass through phases of super-Eddington growth; this is commonly assumed to be possible in the tidal disruption literature but the exact accretion physics remains uncertain.
 
In order for runaway growth to occur, the star cluster must be dense, with a SBH-star tidal interaction rate $\dot{N}_{\bullet \star} \gtrsim 10^{-8}~{\rm yr}$.  This corresponds to physical (cluster core) densities $n_{\rm c} \gtrsim 10^7 ~{\rm pc}^{-3}$.  From a theoretical perspective, such densities are achievable in a NSC that enters a state of core collapse.  Although we consider additional physical processes such as dynamical friction and the finite time required for binary burning, we are in general agreement with the argument of \citet{MilDav12} that primordial binaries should not be able to halt core collapse in clusters with $\sigma \gtrsim 40~{\rm km~s}^{-1}$.  

The scenario presented in this paper differs from analogous ``collisional runaway'' scenarios in the literature by its delayed nature.  Previous runaway scenarios typically invoke much higher densities $\rho_{\rm c}$ so that either a supermassive star can be formed on a timescale short compared to the stellar evolution time $\sim 10^6 ~{\rm yr}$ \citep{Katz+15}, or so that runaway GW capture can occur in a dark subcluster of compact remnants without termination due to ejection \citep{Davies+11}.  Unlike these scenarios, ours unfolds over longer timescales and is therefore not as promising a candidate for production of high-$z$ quasars.  However, the central stellar densities required by our scenario appear achievable in many current low-$z$ NSCs, making a process like the one described in this paper relevant for the evolution of small galactic nuclei.  More specifically, we expect massive black holes to be largely absent from NSCs with $\sigma \lesssim 40~{\rm km~s}^{-1}$, and generally present in larger NSCs.

The general physical considerations outlined above appear to be consistent with current observational data on NSCs.  Firstly, the distribution of dynamically measured SMBH masses and mass upper limits is in good agreement with the predicted transition $\sigma$ from the \citet{MilDav12} argument, and SMBH masses above this transition $\sigma$ agree with our predicted saturation mass $M_{\rm sat}$.  In significantly larger galaxies, $M_\bullet \gg M_{\rm sat}$, implying SMBH growth in this regime is dominated by non-tidal processes, as is expected from the Soltan-Paczynski argument.  Secondly, we analyze the NSC sample of \citet{GeoBok14} and find that essentially no NSCs with $\sigma \lesssim 40~{\rm km~s}^{-1}$ are dense enough to be unstable to a tidal runaway.  
For a sufficiently large velocity dispersion system, we predict either a SMBH grown up to the saturation mass via the tidal processes we have discussed, or one which has passed this point and has grown to still larger mass via large scale gas accretion.

In this paper, we have shown that tidal capture and disruption of stars will inevitably grow stellar mass black holes into IMBHs or SMBHs in less than a Hubble time provided the host star cluster is sufficiently dense.  Although large SMBHs in the universe ultimately grow via other mechanisms \citep{Soltan82}, the processes outlined in this paper appear sufficient to explain the bottom end of the SMBH mass function.  The critical density required to initiate runaway growth {\it is not observed to occur} in the smallest NSCs, as we expect from simple theoretical considerations first outlined in \citet{MilDav12}.  However, larger NSCs often exceed this central density threshold, and many of these NSCs likely possess unresolved massive black holes.  The predictions of our model would be improved by future dynamical work to determine the evolution of tight stellar clusters formed by tidal capture, as well as hydrodynamical simulations of TDEs that robustly determine the fraction of stellar mass accreted.  On the observational side, X-ray observations \citep[for example]{Pandya+16} of the densest NSCs may be useful for determining the present-day rate of black hole growth in these dynamically rich systems.

\section*{Acknowledgments}

We thank Torsten B{\"o}ker, Iskren Georgiev, Dong Lai, Nathan Leigh, Cole Miller, and Scott Tremaine for useful discussions.  We thank Jenny Greene for discussions and for sharing several maser SMBH mass estimates.  We also thank Seppo Mikkola for feedback and for providing his \textsc{AR-Chain} code. NCS acknowledges support through NASA from Einstein Postdoctoral Fellowship Award Number PF5-160145, through NSF grant AST-1410950, and from the Alfred P. Sloan Foundation to Brian Metzger.  AHWK acknowledges support from NASA through Hubble Fellowship grant HST-HF-51323.01-A awarded by the Space Telescope Science Institute, which is operated by the Association of Universities for Research in Astronomy, Inc., for NASA, under contract NAS 5-26555.    This work was aided by the hospitality of the Aspen Center for Physics.

\appendix 

\section{Distributions of NSC Parameters}
\label{app:data}
In this paper, we frequently compare simple theoretical predictions to distributions of observed properties in a large sample of real NSCs.  Our sample includes 22 NSCs from \citet{Boker+04}, 51 NSCs from \citet{Cote+06}, and 151 NSCs from \citet{GeoBok14}.  After discarding 10 NSCs that overlap between these samples we are left with 214 unique NSCs in our broader sample, 207 of which have been fit to three-parameter King models \citep{King66} for which the total mass $M_{\rm tot}$, half-mass radius $r_{\rm h}$, and concentration parameter $C$ are measured.  Because of computational limitations, only a small number of values are considered for the concentration parameter: $C \in \{ 5, 15, 30, 100\}$.  For this reason ``cluster-averaged'' quantities such as $\bar{\rho}$ should be considered much more reliable than ``cluster-central'' quantities such as $\rho_{\rm c}$, and for many of these clusters we only have lower limits on their concentration and central density.

In Figs. \ref{fig:NSCData} and \ref{fig:NSCData2} we fit 2D Gaussians to various combinations of derived and measured parameters.  In the hope that these may be of use to the reader, we provide our best fit relations in Table \ref{tab:distributions}.

\begin{table}
\caption{The 2D Gaussian fits for joint distributions of $\log_{10}X$ and $\log{10}(\sigma~{\rm km}^{-1}~{\rm s})$.  Here $X$ represents a number of different measured and derived parameters of interest.  The covariance matrix for the data set has two eigenvectors whose eigenvalues are $\Lambda_1$ and $\Lambda_2$.  The first eigenvector makes an angle $\theta$ with the positive $\sigma$-axis.}\label{tab:distributions}
\begin{tabular}{| c | c | c | c | c |}
$X$  &  $\langle \log_{10} X \rangle$ & $\Lambda_1$  & $\Lambda_2$ & $\theta$   \\
\hline\hline
$M_{\rm tot}/M_\odot$ &   6.44 & $0.702$ & $0.0318$ & $1.19$ \\
$\bar{\rho}/(M_\odot/{\rm pc}^3)$ &  4.26 & $1.42$  & $0.0628$ & $1.36$ \\
$\bar{t}_{\rm r}/{\rm yr}$ & 9.83 & $0.724$ & $0.115$ & $1.45$ \\
$r_{\rm h}/{\rm pc}$ &  0.519 & $0.187$ & $0.119$ & $-1.31$ \\
\hline
$\rho_{\rm c}/(M_\odot/{\rm pc}^3)$ & 7.61 & $2.85$ & $0.0647$ & $1.43$ \\
$t_{\rm r}(0)/{\rm yr}$ &  6.48 & $1.46$ & $0.122$ & $-1.54$\\
$r_{\rm c}/{\rm pc}$ & -1.16 & $0.439$ & $0.105$ & $-1.33$ \\
\end{tabular}	
\end{table}

\section {Gravitational Wave Capture of Compact Objects} 
\label{sec:GW}
In a core-collapsed cluster, the densities of compact objects will be much higher than those of main sequence stars.  By compact objects, we refer only to neutron stars and other stellar-mass black holes, as white dwarfs are large enough to be tidally disrupted by black holes up to masses of $M_{\rm BH} \approx 10^5 M_{\odot}$.  Denser objects will emit significant energy in gravitational radiation during close encounters.  An unbound pair of compact objects can thus become bound via a gravitational brehmsstralung process; the maximum impact parameter for this to occur is \citep{Peters64, QuiSha87}
\begin{align}
b_{\rm GW} &= \left( \frac{340\pi}{3} \right)^{1/7} \frac{GM_{\rm tot}}{c^2} \frac{\eta^{1/7}}{(v_\infty/c)^{9/7}} \\
& = 7.7R_{\odot}~ \frac{M_{\rm tot}}{20M_{\odot}} \left( \frac{\eta}{1/4} \right)^{1/7} \left( \frac{\sigma}{40~{\rm km~s}^{-1}} \right)^{-9/7} \notag,
\end{align}
corresponding to a maximum pericenter distance of 
\begin{align}
R_{\rm p, GW} & = \left( \frac{85 \pi}{6\sqrt{2}} \right)^{2/7} \frac{GM_{\rm tot}}{c^2} \frac{\eta^{2/7}}{(v_\infty/c)^{4/7}} \\
& = 0.013R_{\odot}~ \frac{M_{\rm tot}}{20M_{\odot}} \left( \frac{\eta}{1/4} \right)^{2/7} \left( \frac{\sigma}{40~{\rm km~s}^{-1}} \right)^{-4/7}, \notag
\end{align}
where the symmetric mass ratio $\eta = M_1 M_2/M_{\rm tot}^2$ and $v_{\infty}$ is the velocity of the two compact objects at infinity.  In the second line of this equation we have set $v_{\infty}=\sigma$.  Following this capture event, the compact object binary will evolve from an extremely high-eccentricity orbit under the influence of gravitational radiation.  

The very small cross-section for GW capture disfavors this SBH growth mechanism relative to tidal interactions with stars.  GW capture is further disfavored by the large recoil velocities the SBHs will receive due to anisotropic emission of GWs during their merger.  These velocities are typically $\gtrsim 100 ~{\rm km~s}^{-1}$, but with a long tail going up to $\sim 5000~{\rm km~s}^{-1}$ \citep{Lousto12}.  This is large enough to eject the merged SBH from not just its NSC, but frequently even its host galaxy and halo.  

\section{Tidal Disruption of Stars} 
\label{sec:TDE}
At intermediate impact parameters, stars passing close to stellar mass black holes will be tidally disrupted.  Pericenters vulnerable to tidal disruption will lie interior to the tidal radius,
\begin{equation}
R_{\rm t} = R_* \left( \frac{M_{\bullet}}{M_*} \right)^{1/3}.
\end{equation}
Following disruption, the stellar debris will follow ballistic trajectories with a frozen-in spread of specific energy \citep{Rees88, Stone+13}, 
\begin{equation}
\Delta \epsilon = \frac{GM_{\bullet} R_*}{R_{t}^2}.
\end{equation}
Generally, $\Delta \epsilon \gg \sigma^2$, meaning that half the star will be unbound and fly off to infinity, while the other half will eventually return to the SBH on highly eccentric orbits.  These bound debris streams will collisionally shock each other and circularize into a super-Eddington accretion disk after order unity pericenter returns, although the details of circularization are unclear: although they are likely mediated by GR precession around SMBHs \citep{Hayasa+13, Hayasa+15}, near stellar mass BHs the tidal radius is very non-relativistic and purely hydrodynamic effects may prevail \citep{Guillo+14, Shioka+15}.  Regardless, a SBH will likely accrete $\approx 50 \%$ of the tidally disrupted star.  

Recently, \citet{MetSto15} suggested that the true accretion fraction may be significantly lower than $50\%$ because of the large energy hierarchy in the problem: $\Delta \epsilon \gg \sigma^2$, so the absorption of even a small fraction of the accretion luminosity will suffice to unbind a majority of the stellar debris bound to the black hole.  If this argument is correct, it would drastically reduce the ability of TDEs to contribute to SBH mass growth.  However, the analytic argument of \citet{MetSto15} was designed to apply to SMBH TDEs, where the energy hierarchy is more than an order of magnitude stronger than it is for SBH TDEs.  Realistic numerical simulations capable of testing this argument have not yet been run for SMBH TDEs, but at lower mass ratios ($M_\bullet = 10^3 M_\odot$) they have, and there is no evidence for the ejection of the bound debris \citep{Guillo+14}.

\section{Tidal Capture of Stars} 
\label{sec:TC}
\begin{figure}
\includegraphics[width=85mm]{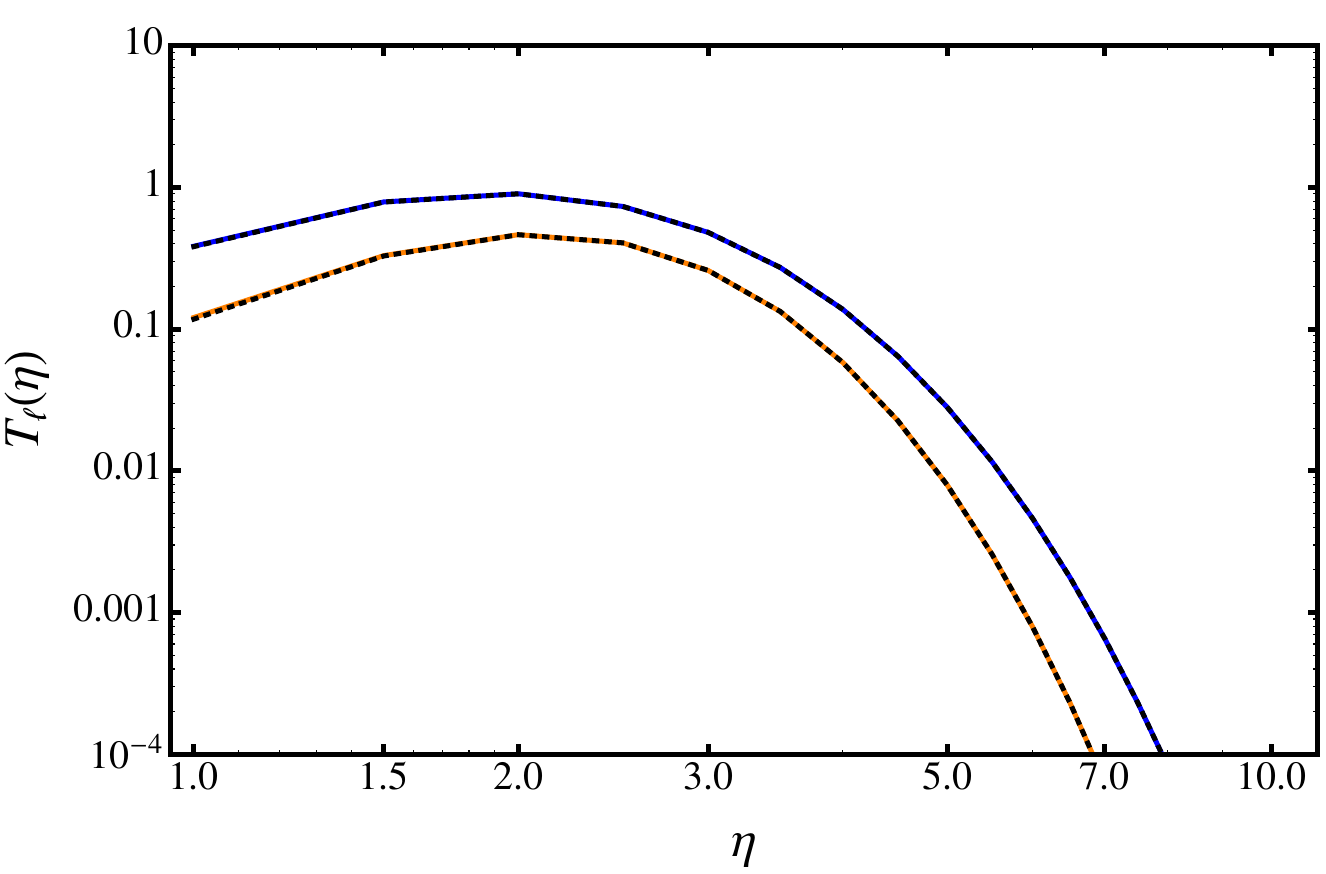}
\caption{The dimensionless tidal coupling constants, $T_{\ell}(\eta)$, for both $\ell=2$ (blue) and $\ell=3$ (orange) modes in a $n_{\rm poly}=3/2$ polytropic star.  Calculation of $T_2$ and $T_3$ involves a summation over the lowest-order p-modes, but in general it is the f-mode that dominates.  These curves are independent of perturber mass ratio.}
\label{TlCurve}
\end{figure}

\begin{figure}
\includegraphics[width=85mm]{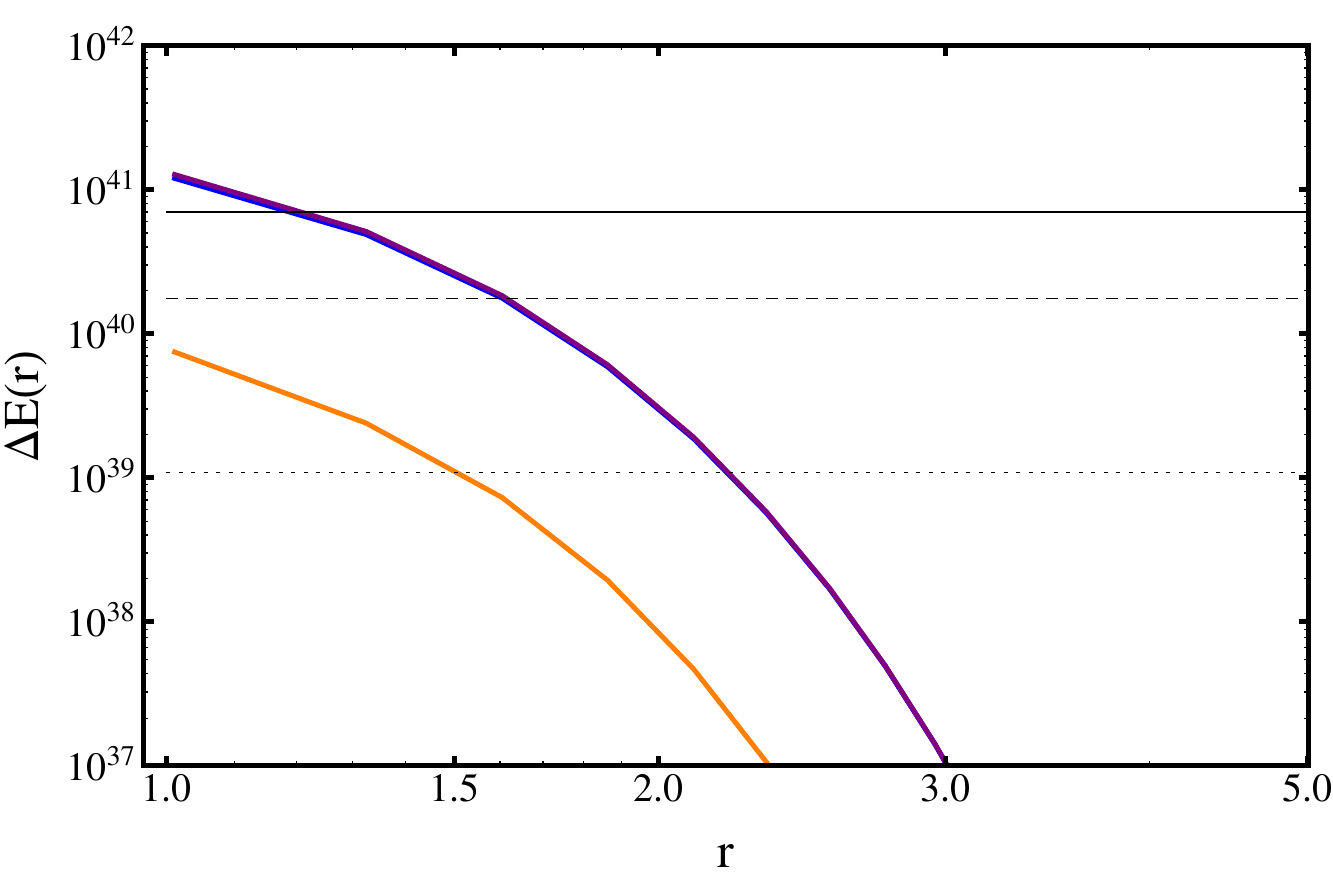}
\caption{The total energy (units of J) deposited into mechanical oscillations of a solar-type, $n_{\rm poly}=3/2$ star perturbed by a $10M_{\odot}$ BH, as a function of the pericenter radius $r$ in units of tidal radii.  The blue curve shows the contribution of $\ell=2$ modes, the orange curve the contribution of $\ell=3$ modes, and the purple curve (overlapping with blue) their combined effect.  The horizontal black lines show the typical energy at infinity of clusters with $\sigma=80~{\rm km~s}^{-1}$ (solid), $\sigma=40~{\rm km~s}^{-1}$ (dashed), and $\sigma=10~{\rm km~s}^{-1}$ (dotted).}
\label{DeltaECurve}
\end{figure}

\begin{figure}
\includegraphics[width=85mm]{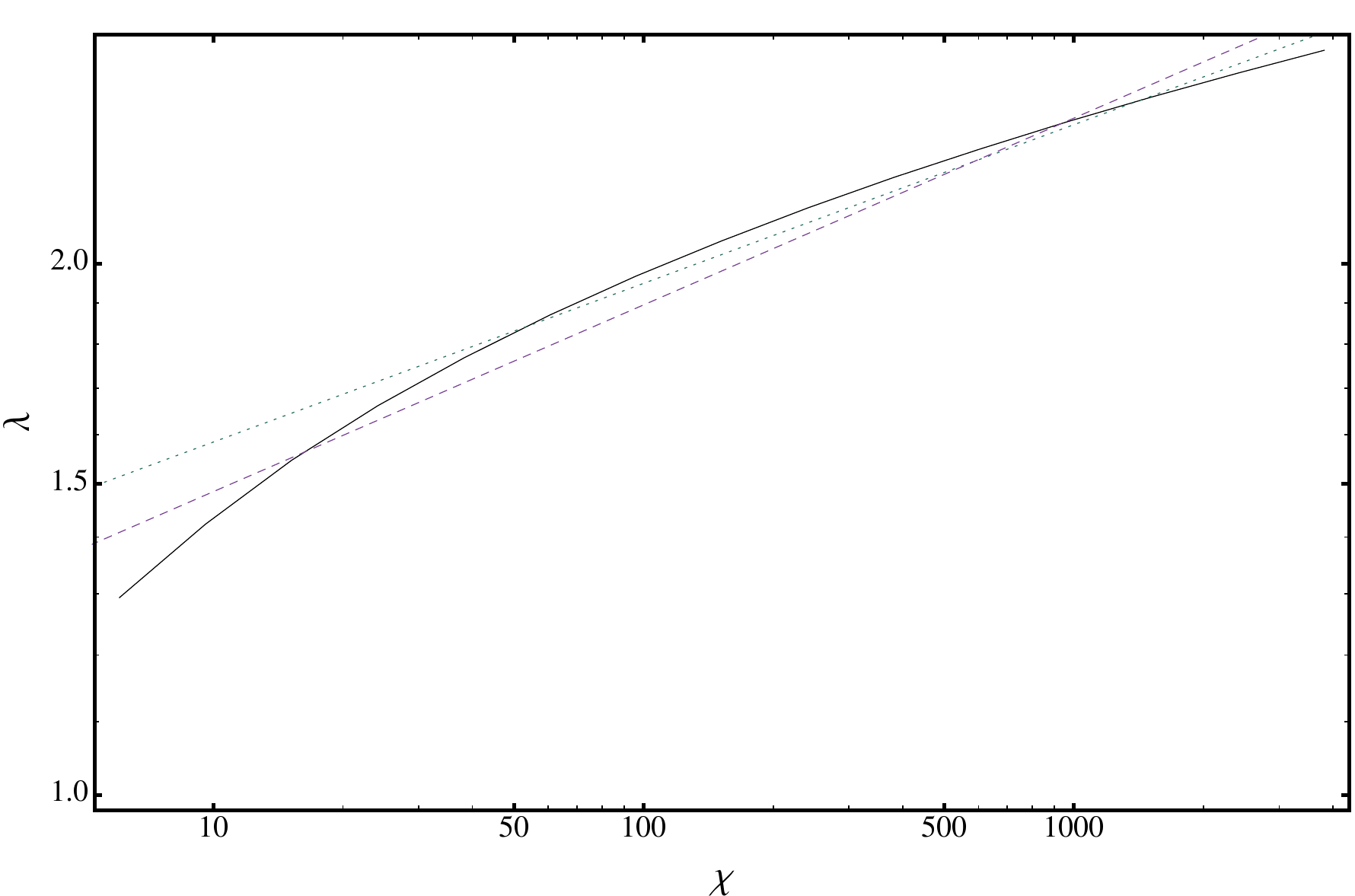}
\caption{The $\lambda$ parameter required for tidal capture shown as a function of $\chi \equiv 2GM_\star/(R_\star \sigma^2)$.  The black line is an exact numerical solution, and we also show power law best fits for $\chi < 0.05$ (dotted green line) and $\chi < 0.15$ (dashed purple line).  The former, fiducial curve is given in the text as Eq. \ref{eq:chiFit}, while the alternate one (which may extend too far into the nonlinear tides regime, beyond which our treatment of mode excitation breaks down) is $\lambda = 1.165 \chi^{0.1058}$. }
\label{chiFit}
\end{figure}

At larger separations, tidal excitation of modes can bind a SBH to a star initially on a mildly hyperbolic orbit.  The excess orbital energy is deposited into the mechanical oscillation of normal modes in the star, putting the two onto an eccentric bound orbit.  In order for successful capture to occur the energy deposited into modes, $\Delta E$, must exceed the hyperbolic orbital energy $E_{\rm orb} \approx v_{\infty}^2/2$.  If we decompose the modes into spherical harmonics $\{\ell, m\}$ with radial mode numbers $n$, we can follow \citet{PreTeu77} to write
\begin{equation}
\Delta E_{\ell} = \frac{GM_*^2}{R_*} \left( \frac{M_\bullet}{M_*} \right)^2 \sum_{\ell=2, 3, ...}^{\infty} \left( \frac{R_*}{R_{\rm p}} \right)^{2\ell + 2} T_{\ell} (R_{\rm p}).
\end{equation}
Here the dimensionless mode-orbit coupling constant
\begin{equation}
T_{\ell}(R_{\rm p}) = 2\pi^2 \sum_n^\infty | Q_{n \ell} |^2 \sum_{m=-\ell}^{\ell} |K_{n\ell m} |^2
\end{equation}
is broken down into two distinct components: a dimensionless overlap integral $Q_{n\ell}$ measuring normal modes inside a given stellar structure model, and a dimensionless tidal coupling integral $K_{n \ell m}$.  We calculate the second of these directly, using 
\begin{align}
K_{n \ell m} = &\frac{W_{\ell m}}{2\pi} \int_{-\infty}^{\infty} {\rm d}t \left( \frac{R_{\rm p}}{R(t)} \right)^{\ell + 1} e^{i(\omega_{n}t+m\phi)} \\
W_{\ell m} = &(-1)^{\frac{\ell + m}{2}} \left( \frac{4\pi}{2\ell + 1}(\ell-m)! (\ell + m)! \right)^{1/2} \\
&\times \left(2^\ell \left( \frac{\ell +m}{2} \right)!  \left( \frac{\ell -m}{2} \right)!  \right)^{-1}. \notag
\end{align}
In this coupling integral $\phi$ is orbital phase and $\omega_n$ is the mode frequency.  Values of $Q_{n \ell}$ have been tabulated in other papers \citep{LeeOst86}, and we use these tabulations for the low-$\ell$ modes which dominate energy deposition.  Sample tidal coupling curves $T_2(\eta)$ and $T_3(\eta)$ are shown in Fig. \ref{TlCurve}.  In practice, the vast majority of energy deposited in typical tidal capture events comes from $\ell=2, n=0$ oscillations.  Numerically, one can calculate $\Delta E_{\ell}$ to find the critical pericenter $R_{\rm TC}$ that results in tidal capture; if we parametrize $R_{\rm TC} \equiv \lambda R_{\rm t}$ we find that $\lambda \approx 2$ in a cluster with $\sigma_{40} = 1$.  This is shown in more detail in Fig. \ref{DeltaECurve}, and quantified in Fig. \ref{chiFit}, where we show how $\lambda$ scales with the dimensionless number $\chi \equiv 2GM_\star/(R_\star \sigma^2)$, the squared ratio of the star's surface escape velocity to the cluster $\sigma$.  Included in this plot are power-law best fits to exact solutions; if we cut off the best fit at $1/\chi \sim \Delta E/(GM_\star^2/R_\star) < 0.05$ (an approximation for where the linear tides approximation is valid) we have 
\begin{equation}
\lambda = 1.288 \chi^{0.0899}. \label{eq:chiFit}
\end{equation}

\section{N-body simulations}\label{sec:simulations}
For the numerical tests presented in \S~\ref{sec:numerics}, we used two different integrators: one for single-body integrations and one for few-body integrations. The single-body code is a simple Hermite integrator with a fixed time step, whereas the few-body code is a modified version of the algorithmic chain integrator \textsc{AR-Chain} developed by \citet{Mikkola06}. The latter uses algorithmic chain regularization for high-precision integration of few-body dynamics, and is capable of handling velocity-dependent forces efficiently. It includes relativistic post-Newtonian terms up to order PN2.5 \citep{Mikkola08}.

Both codes integrate the orbits of SBHs in the gravitational field of a background NSC. For computational convenience, this star cluster is approximated by a Plummer sphere with total mass, $M_{NSC}$, and Plummer scale radius, $a$. Following, e.g., \citet{Heggie03}, its mass density profile can be written as
\begin{equation}
\rho(r) = \frac{3M_{NSC}}{4\pi a^3}\left(1+\frac{r^2}{a^2} \right)^{-5/2},
\end{equation}
having a core of size $r_c = a/\sqrt{2}$ with a central density of $\rho_0  = 3M_{NSC}/(4\pi a^3)$. Its half-mass radius is given by $r_h \approx 1.305 a$. The velocity dispersion at radius $r$ can be written as
\begin{equation}
\sigma (r) = \sqrt{\frac{GM_{NSC}}{6a}\left( 1 + \frac{r^2}{a^2}\right)^{-1/2}},
\end{equation}
with a central velocity dispersion of $\sigma_0 = \sqrt{GM_{NSC}/6a}$. These quantities, density and velocity dispersion, are of key importance for the runaway time scale of the SBHs.
We relate these three-dimensional quantities to observables by defining the NSC's (projected) effective radius as $R_{eff} = a/\sqrt{2^{2/3}-1}$, and the velocity dispersion at this radius to be
\begin{equation}
\sigma_{eff} = 22.42\left(\frac{M_{NSC}}{10^6\,\mbox{M}_\odot}\right)^{1/2}\left(\frac{1\,\mbox{pc}}{R_{eff}}\right)^{1/2}\,\mbox{km\,s}^{-1}
\end{equation}

\subsection{Phase-space diffusion}
Weak encounters with background stars will let the SBHs diffuse through phase space while they are orbiting within the gravitational potential of the NSC. The diffusion can be expressed as change in velocity of an SBH by $\Delta \vec{v}$ per unit time. We can split this change into a component along the direction of motion of the SBH, and one perpendicular to that. Following \citet{BinTre08}, the diffusion coefficients can be expressed as
\begin{eqnarray}
D[\Delta v_\parallel] & = & -\frac{4\pi G^2\rho(r)M_\bullet\ln\Lambda}{\sigma^2}f(\chi),\label{eq:df}\\
D[(\Delta v_\parallel)^2] & = & \frac{4\sqrt{2}\pi G^2\rho(r)M_\bullet\ln\Lambda}{\sigma}\frac{f(\chi)}{\chi},\\
D[(\Delta \vec{v}_\bot)^2] & = & \frac{4\sqrt{2}\pi G^2\rho(r)M_\bullet\ln\Lambda}{\sigma}\left[\frac{\mbox{erf}(\chi)-f(\chi)}{\chi}\right],
\end{eqnarray}
where $\Delta v_\parallel \equiv \Delta \vec{v}\cdot\vec{v}/v$ is the velocity change in direction of motion, and $\Delta \vec{v}_\bot \equiv \Delta \vec{v} - \Delta v_\parallel \cdot\vec{v}/v$ is the velocity change perpendicular to the direction of motion. Here, $M_\bullet$ is the mass of the black hole, and $\chi = \frac{v}{\sqrt{2}\sigma(r)}$. The function $f(\chi)$ is given by
\begin{equation}
f(\chi) = \frac{1}{2\chi^2}\left(\mbox{erf}(\chi)-\frac{2\chi}{\sqrt{\pi}}\exp\left(-\chi^2\right)\right).
\end{equation}
We approximate the factor $\Lambda$ in the Coulomb logarithm as
\begin{equation}
\Lambda = \left(\frac{M_{NSC}}{M_\bullet}\right)\left(\frac{r}{r_h}\right).
\end{equation}
We can identify Eq.~\ref{eq:df} as the dynamical friction term, that is, if we assumed $D[(\Delta v_\parallel)^2]  = D[(\Delta \vec{v}_\bot)^2]  = 0$, we would get Chandrasekhar's dynamical friction formula. The second term introduces a variance of the friction term, and even allows the SBHs to be accelerated when the velocity of a SBH gets sufficiently small. The third term introduces a change in velocity perpendicular to the direction of motion of the SBH. It is a randomly oriented vector, and hence causes the SBHs to execute a random walk in phase space. The last two terms will establish that the SBHs are ultimately in energy equipartition with the background stars.
The velocity changes $\Delta v_\parallel$ and $\Delta\vec{v}_\bot$ per unit time $\Delta t$ can be computed with the above equations. Both changes are normal distributed, where the mean, $\mu$, and the variance, $\Sigma$, of the distributions are given by
\begin{eqnarray}
\mu_\parallel &=& D[\Delta v_\parallel]\Delta t,\\
\Sigma_\parallel &=& D[(\Delta v_\parallel)^2]\Delta t,\\
\mu_\bot &=& 0,\\
\Sigma_\bot &=& D[(\Delta \vec{v}_\bot)^2]\Delta t.
\end{eqnarray}
We compute the diffusion coefficients for each black hole at each time step, and modify its velocity on a Monte Carlo basis. For each time step we draw a random orientation before adding the perpendicular velocity change to the respective SBH. Hence, the SBH's modified velocity, $v_f$, is computed using
\begin{eqnarray}
\vec{v}_f &=& \vec{v}_0 + \Delta v_\parallel \hat{v}_\parallel + \Delta v_\bot \hat{v}_\bot,\\
\Delta v_\parallel &=& \mathcal{N}(\mu_\parallel, \Sigma_\parallel),\\
\Delta v_\bot &=& \mathcal{N}(\mu_\bot, \Sigma_\bot).
\end{eqnarray}
The change of energy, $\mbox{d}E_{BH}$, of the orbiting black hole due to phase-space diffusion is given back to the stellar background potential, with $\mbox{d}E = -\mbox{d}E_{BH}$. As a consequence of this energy transfer, inspiralling black holes will cause an expansion of the NSC. For this purpose we calculate the change in potential energy, $\mbox{d}W$, of the stellar system using
\begin{eqnarray}
E &=& T + W = \frac{1}{2}W,\\
\mbox{d}W &=& -2\,\mbox{d}E_{BH},
\end{eqnarray}
where we made use of the virial theorem $2T+W =0$. With this change in potential energy we can calculate a new radius for the stellar background potential at each integration step. For the Plummer sphere the new scale radius can be calculated as
\begin{equation}
a_{new} = a\left(1+\frac{32a\,\mbox{d}W}{3\pi GM_{NSC}^2}\right)^{-1}.
\end{equation}

\subsection{Tidal captures and tidal disruptions}
For each SBH we compute the probability of capturing or disrupting one or many background stars during each time step. According to Eq.~\ref{eq:TCRate}, the tidal-capture rates are estimated as
\begin{equation}
\dot{N}_{\bullet \star} = n_{\star}\,\Sigma\,v_{rel},
\end{equation}
where the number density of stars is $n_\star$ = $\rho(r)/M_\star$ with a mean stellar mass of 0.45\,M$_\odot$. The tidal-capture cross section, $\Sigma$, is given by
\begin{equation}
\Sigma = \pi\,R_{t}^2\left(1+\frac{2GM_{tot}}{R_t\,v^2_{rel}} \right),
\end{equation}
where $M_{tot}$ is the total mass of SBH plus star, and $v_{rel}$ is the relative velocity between the two (which we approximate as the local velocity dispersion $\sigma$). The tidal radius is given by $R_t = r_\star\left(M_\bullet/M_\star\right)^{1/3}$, with $r_\star$ being the mean stellar radius approximated as $r_\star \approx r_\odot\,(M_\star/\mbox{M}_\odot)^{0.8}$.
From these rates we calculate tidal capture probabilities by assuming they are constant throughout one time step. Since our time steps are small fractions of the NSC's crossing time, this assumption is a good approximation.

Using this tidal capture probabilities, we let the SBHs grow on a Monte Carlo basis. We distinguish between tidal capture and disruption events, which have related cross sections, by assuming that a fraction of the capture events are disruptions. Each tidal capture results in the respective SBH gaining one mean stellar mass, $M_\star$, whereas disruptions increase the SBH mass by $0.5\,M_\star$ (see \S 3).

However, there remains a notable shortcoming of this numerical approach:  the Plummer sphere has a fixed ratio of core radius to half-mass radius ($r_c/r_h\approx 0.54$); realistic clusters exhibit wide variance in this ratio, and the large central densities of highly concentrated clusters are more favorable to tidal capture runaways.  In this sense our simulations are conservative with regard to the onset of runaway SBH growth, and we are currently performing analogous few-body simulations using the potential-density pair of \citet{StoOst15} to generalize these results.

\subsection{BH-BH mergers and dynamical ejections}
The code \textsc{AR-Chain} includes PN terms up to order 2.5. The SBHs can therefore merge via gravitational wave emission. We include gravitational wave recoils following the prescription outlined in \citet{Kulier15}, which is based on the fitting formula by \citet{Lousto12}. To save computational time, we assume that a merger will be inevitable when the separation between two SBHs gets smaller than 10\,000 Schwarzschild radii. At the moment of the merger, we assume that the spin vectors of the two SBHs are randomly aligned. This results in kick velocities of up to several thousand km\,s$^{-1}$, with a median kick of $\approx 290$\,km\,s$^{-1}$. Since our simulations focus on NSC with relatively low escape velocities, this implies that a majority of the merging SBHs escape from the NSCs.

Black holes can also eject each other via strong three-body interactions. We remove SBHs from the simulations once they move beyond 1\,kpc from the NSC, assuming that it will take them more than a Hubble time to find their way back into the center of the host galaxy.

\subsection{Set of simulations}
The simulation results presented in \S 4 are based on a set of 9 runs with the single-body code and 9 runs with the few-body code. The characteristics of the NSCs in both sets are the same, but while the one-body simulations show us the evolution of a single SBH in these NSCs, the few-body code follows 10 SBHs as they grow and interact.

We use the \textsc{McLuster} code\footnote{https://github.com/ahwkuepper/mcluster} to generate initial conditions for these simulations \citep{Kupper11}. The 10 SBH follow a simple mass function with slope $M_\bullet^{-1}$ between 5\,M$_\odot$ and 20\,M$_\odot$, resulting in the most massive SBH having 20\,M$_\odot$ and the next-most massive SBH having a mass of 11\,M$_\odot$. The single-body simulations are using only the most-massive SBH of these initial conditions.

We make five simulations with a fixed NSC mass of 10$^6$\,M$_\odot$ and effective radii varying between 0.1\,pc and 2\,pc. Another four simulations have a fixed effective radius of 0.2\,pc and NSC masses between $2.5\cdot10^5$\,M$_\odot$ and $2\cdot10^6$\,M$_\odot$. The SBHs are set up randomly within these NSCs with velocities typical for their initial NSC positions. The results are described in \S 4 and summarized in Fig.~\ref{fig:fewBody}.

\end{document}